\documentclass{article}
\usepackage{PRIMEarxiv}

\usepackage[utf8]{inputenc}
\usepackage[T1]{fontenc}
\usepackage{float}
\usepackage{cprotect}
\usepackage{listings}
\usepackage{xcolor}
\usepackage{dirtree}
\usepackage{mathtools}
\usepackage{comment}
\usepackage[labelfont=bf,font=small,skip=5pt]{caption}
\usepackage[outdir=figures/]{epstopdf}
\usepackage{titlesec}
\usepackage{subcaption}
\usepackage{amssymb}
\usepackage{amsmath}

\usepackage{hyperref}

\usepackage{etoolbox}
\apptocmd{\sloppy}{\hbadness 10000\relax}{}{}

\definecolor{codegreen}{rgb}{0,0.6,0}
\definecolor{codegray}{rgb}{0.5,0.5,0.5}
\definecolor{codepurple}{rgb}{0.58,0,0.82}
\definecolor{backcolour}{rgb}{0.95,0.95,0.92}

\title{Accelerated Patient-Specific Calibration via Differentiable Hemodynamics Simulations
}

\author{
  Diego Renner \\
  Seminar for Applied Mathematics \\
  ETH Zurich\\
  \texttt{drenner.business@gmail.com} \\
     \And
  Georgios Kissas \\
  AI Center \\
  ETH Zurich \\
  \texttt{gkissas@ai.ethz.ch} \\
}
\usepackage{kantlipsum}

\begin{document}

\maketitle

\begin{abstract}
One of the goals of personalized medicine is to tailor diagnostics to individual patients. Diagnostics are performed in practice by measuring quantities, called biomarkers, that indicate the existence and progress of a disease. In common cardiovascular diseases, such as hypertension, biomarkers that are closely related to the clinical representation of a patient can be predicted using computational models. Personalizing computational models translates to considering patient-specific flow conditions, for example, the compliance of blood vessels that cannot be a priori known and quantities such as the patient geometry that can be measured using imaging. Therefore, a patient is identified by a set of measurable and nonmeasurable parameters needed to well-define a computational model; else, the computational model is not personalized, meaning it is prone to large prediction errors. Therefore, to personalize a computational model, sufficient information needs to be extracted from the data. The current methods by which this is done are either inefficient, due to relying on slow-converging optimization methods, or hard to interpret, due to using `black box` deep-learning algorithms. We propose a personalized diagnostic procedure based on a differentiable 0D-1D Navier-Stokes reduced order model solver and fast parameter inference methods that take advantage of gradients through the solver. By providing a faster method for performing parameter inference and sensitivity analysis through differentiability while maintaining the interpretability of well-understood mathematical models and numerical methods, the best of both worlds is combined. The performance of the proposed solver is validated against a well-established process on different geometries, and different parameter inference processes are successfully performed. 
\end{abstract}

\section{Introduction}

The cardiovascular system and cardiovascular disease are prime targets for personalized medicine approaches. The significance of cardiovascular disease and its research is due to it being the most prevalent cause of death. Worldwide, it led to 17.3 million deaths per year as of 2015. This number is projected to increase by 2030 to more than 23.6 million. \cite{update2015heart}. In Europe, cardiovascular disease was responsible for close to one third (32.7\%) of all deaths in the year 2020 \cite{Coelho2020}.

Taking into account the characteristics of an individual patient and customizing their treatment according to which sub-population they fall into, personalized medicine aims to improve clinical care \cite{ashley2016towards,national2011toward,giardino2017role}. To succeed in the personalization of a treatment, disease-specific quantities (biomarkers) must be accurately measured or predicted \cite{burke1998integrating}. Biomarkers can have a varying degree of correlation with the existence, progression, and outcome of a disease. For example, local vascular pressure is strongly correlated with hypertension. Biomarkers such as pulsatility and resistivity indices extracted from ultrasound or Magnetic Resonance Imaging \cite{markl20124d} establish an implicit relation between vessel flow rate and pressure, but are not accurate enough for standard clinical practice \cite{everett2012beyond}. Invasive techniques, for example, the introduction of pressure catheters in blood vessels, also exist, but are associated with the risk of microsurgery and, in cases such as pregnancy, unethical \cite{kett2002adverse}. Biomarkers strongly correlated with the outcome of diseases such as hypertension, aneurysm development, and stenosis can be computed using computational models. 

Such computational models aim to simulate the individual physiology of a patient to predict the aforementioned "hidden" biomarkers. This is especially useful when computed biomarkers can be connected by first principles to the outcome of a disease, such as wall shear stress to the development of the aneurysm. However, simulating such quantities accurately depends heavily on a high number of personalized parameters, leading to a similar problem as described initially: the need to access difficult-to-measure or unobtainable data for a specific patient. An attempt to avoid the problem of having to know these values for each individual is to use population-averaged means. However, this defeats the purpose of personalized medicine. For calibrating a solver to specific conditions, roughly two families of approaches have been developed. The first approach considers the inverse problem of inferring parameters from data in a probabilistic manner \cite{lingsch2024fuse, kissas2023towards} by pretraining a deep learning model on a patient data set. For a new patient, the parameters are first inferred, and then the computational model is solved for the inferred parameters to obtain predictions of the biomarker and uncertainty estimates \cite{kissas2022feasibility,richter2024bayesian, kissas2023towards}. The second approach considers a methodology in which the computational model parameters are sampled from a prior distribution; they are used as input to a solver, and then the likelihood between the solver predictions and the measured data is considered. Both pipelines fall under Simulation Based Inference (SBI) \cite{cranmer2020frontier}.

Despite the success of the first approach in the monitoring of cardiovascular health \cite{wehenkel2023simulation}, it is prone to out-of-distribution errors and misspecification problems \cite{wehenkel2024addressing}. The second approach does not consider amortization and therefore must be run from scratch for each individual patient, which is shown to be a very time-consuming process when done naively \cite{kissas2022feasibility}. This is in part because derivatives are hard to determine for attempted simulations, so the algorithms of choice for solving the inverse problem are gradient-free. Moreover, because models are very sensitive to their parameters and there are ambiguous solutions \cite{nolte2022inverse,quick2001infinite}, these gradient-free algorithms must be run at very small step sizes \cite{taylor2009patient,tuccio2022parameter,marsden2014optimization,mineroff2019optimization,bozkurt2022patient}. The ability to easily take gradients of one of these models with respect to its parameters would greatly improve the calibration without losing interpretability, as would be the case when using a deep-learning algorithm. Personalized medicine could be specifically aided by using gradients to more efficiently infer parameters or perform sensitivity analysis. Therefore, cardiovascular systems have been well studied, from analytical solutions of simplified models to approximations of complex systems. In order to tackle this complexity of the meaningful task of modeling cardiovascular systems, numerical modeling has become prevalent \cite{formaggia2009multiscale,quarteroni2016geometric,black2020p14,el2018investigating,qureshi2014numerical,reichold2009vascular}. The issue of needing access to hard-to-obtain biomarkers and the pitfalls of the existing simulations that provide such biomarkers hold also in the case of cardiovascular disease.

The goal of this work is to provide a fast, differentiable cardiovascular solver to be used as part of a calibration pipeline. This is a step towards enabling efficient and interpretable model calibration for personalized medicine. In order to be differentiable, the hemodynamic solver is written using the JAX library \cite{jax123}. In addition to differentiability, JAX also offers device-agnostic execution. That is, JAX optimizes the written code to benefit from parallelization and other advantages of modern hardware regardless of being executed on a CPU, GPU, or TPU. The optimization for GPU execution is especially useful since it allows for data parallelization, i.e. batch executions of the code on a single device. This means multiple models for different patients could be optimized simultaneously on a single GPU.


\section{Methods} \label{sec:1dm}
In this section, we discuss different approaches to modeling pulse wave propagation and compare the benefits and argue about the choices made in this manuscript \ref{sec:ma}.

\subsection{Different Modeling Approaches} \label{sec:ma}

3D models are considered for modeling the flow in high resolution, but mostly to predict quantities that require three-dimensional geometric characteristics such as wall shear stresses. However, 3D models are very computationally intensive \cite{arzani2022machine}, especially in the case where moving boundaries are considered. Models that under assumptions reduce the dimensions of the problem offer an alternative for cases where one aims to predict quantities such as mean flow or local vascular pressure, that do not require considering three-dimensional geometric characteristics. One-dimensional reduced-order models are validated in vitro and in vivo and have shown good agreement with full-order simulations. Finally, 0D approaches model the approximate behavior of ensembles of vessels, or organs, and are often used to simulate the outlets of a network modeled in higher dimensions, where it would be too costly to simulate every vessel individually at the increased accuracy. In practice, it is also common to couple such models, choosing the more demanding models for locations in the vascular network that require more detail or that simply are of special interest. This leads to 3D-1D-0D models in which the junctions in the network are represented by 3D models, 1D models are used to simulate the flow within vessels, and 0D models are used to represent the effect of downstream vessels that are excluded from the circulation. The proposed work considers a couple of 1D and 0D models, since the main goal of this work is to produce a fast and differentiable solver.

\subsection{Reduced Order Navier-Stokes Model} \label{sec:sv}
Within a vessel, we consider the conservation of mass, or continuity Equation\cite{anderson2011ebook}:
\begin{equation}
	\frac{\partial \rho (z,r,\phi; t)}{\partial t} + \nabla \cdot \rho (z,r,\phi; t)\mathbf{u}(z,r,\phi; t) = 0, \label{eq:cont}
\end{equation}
and momentum:
\begin{equation}
\begin{aligned}
	\frac{\partial \mathbf{u}(z,r,\phi; t)}{\partial t} + \left( \mathbf{u}(z,r,\phi; t) \cdot \nabla \right) \mathbf{u}(z,r,\phi; t) &- \frac{\mu}{\rho(z,r,\phi; t)} \nabla^2 \mathbf{u}(z,r,\phi; t) = \\
														&- \frac{1}{\rho (z,r,\phi; t)} \nabla P(z,r,\phi; t) + \mathbf{F}(z,r,\phi; t), \label{eq:mass} 
\end{aligned}
\end{equation}
in cylindrical coordinates:
\begin{equation*}
	z, r \in \mathbb{R}, \ t \in \mathbb{R}_{\geq 0}, \ \phi \in [0, 2\pi), \ \nabla := \left[ \frac{\partial}{\partial z}, \frac{1}{r}\frac{\partial}{\partial r}r, \frac{1}{r}\frac{\partial}{\partial \phi}  \right].
\end{equation*}
Here $\rho$, $\mu$ correspond to the blood density and viscosity, respectively, and 
\begin{equation*}
	\mathbf{u}(z,r,\phi; t) := \left[ u_z(z,r,\phi; t), u_r(z,r,\phi; t), u_\phi (z,r,\phi; t) \right]^T
\end{equation*}
is the velocity field, $P$ is the pressure, and $\mathbf{F}$ a body force acting on the fluid. We model a vessel as an axisymmetric tube along the $z$ axis in a cylindrical coordinate system $\left(z,r,\phi\right)$. The Navier-Stokes equations can be reduced under the following assumptions:
\begin{enumerate}
	\item Blood is an incompressible Newtonian fluid, and its properties do not vary over a cross section. \label{it:fld} 
	\item Vessel walls can be displaced in radial direction but not in longitudinal direction, \label{it:displ} 
	\item The length of the vessel is much greater than their radius, \label{it:nlc}
	\item Vessels have elastic compliant walls, \label{it:lec}
\end{enumerate}
In order to make this reduction we first apply the assumption that $\rho$ is constant in the continuity Equationand expand the momentum Equationin it's components
\begin{equation*}
	\frac{\partial u_z}{\partial z} + \frac{1}{r} \frac{\partial (ru_r)}{\partial r} + \frac{1}{r}\frac{\partial u_\phi}{\partial \phi} = 0, 
\end{equation*}
\begin{equation*}
\begin{aligned}
	\frac{\partial u_z}{\partial t} + u_z \frac{\partial u_z}{\partial z} + u_r \frac{\partial u_z}{\partial r} + \frac{u_\phi}{r}\frac{\partial u_z}{\partial \phi} &=
	-\frac{1}{\rho}\frac{\partial P}{\partial z} + \frac{\mu}{\rho} \left( \frac{\partial^2 u_z}{\partial z^2} + \frac{\partial^2 u_z}{\partial r^2} + \frac{1}{r} \frac{\partial u_z}{\partial r} + \frac{1}{r^2}\frac{\partial^2 u_z}{\partial \phi^2} \right), \\
\frac{\partial u_r}{\partial t} + u_z \frac{\partial u_r}{\partial z} + u_r \frac{\partial u_r}{\partial r} + \frac{u_\phi}{r}\frac{\partial u_r}{\partial \phi} -\frac{u_\phi^2}{r} &= 
	-\frac{1}{\rho}\frac{\partial P}{\partial r} + \frac{\mu}{\rho} \left( \frac{\partial^2 u_r}{\partial z^2} + \frac{\partial^2 u_r}{\partial r^2} + \frac{1}{r} \frac{\partial u_r}{\partial r} + \frac{1}{r^2}\frac{\partial^2 u_r}{\partial \phi^2} -\frac{2}{r^2}\frac{\partial u_\phi}{\partial \phi} - \frac{u_r}{r^2} \right), \\
    \frac{\partial u_\phi}{\partial t} + u_z \frac{\partial u_\rho}{\partial z} + u_r \frac{\partial u_\phi}{\partial r} + \frac{u_\phi}{r}\frac{\partial u_\phi}{\partial \phi} -\frac{u_r u_\phi}{r} &= 
	-\frac{1}{\rho}\frac{\partial P}{\partial \phi} + \frac{\mu}{\rho} \left( \frac{\partial^2 u_\phi}{\partial z^2} + \frac{\partial^2 u_\phi}{\partial r^2} + \frac{1}{r} \frac{\partial u_\phi}{\partial r} + \frac{1}{r^2}\frac{\partial^2 u_\phi}{\partial \phi^2} -\frac{2}{r^2}\frac{\partial u_r}{\partial \phi} - \frac{u_\phi}{r^2}\right). \label{eq:momp}
\end{aligned}
\end{equation*}
We dropped the explicit dependence on time and space variables in favor of readability. Since the model is assumed axisymmetric, we set $v_\phi=0$ the system of equations now reads:
\begin{equation}
	\frac{\partial u_z}{\partial z} + \frac{1}{r} \frac{\partial (r u_r)}{\partial r} = 0, \label{eq:cont1}
\end{equation}
\begin{equation*}
	\frac{\partial u_z}{\partial t} + u_z \frac{\partial u_z}{\partial z} + u_r \frac{\partial u_z}{\partial r} = -\frac{1}{\rho} \frac{\partial P}{\partial z} + \frac{\mu}{\rho} \left( \frac{\partial^2 u_z}{\partial z^2} + \frac{\partial^2 u_z}{\partial r^2} + \frac{1}{r} \frac{\partial u_z}{\partial r} \right),
\end{equation*}
\begin{equation}
	\frac{\partial u_r}{\partial t} + u_z \frac{\partial u_r}{\partial z} + u_r \frac{\partial u_r}{\partial r} = -\frac{1}{\rho} \frac{\partial P}{\partial r} + \frac{\mu}{\rho} \left( \frac{\partial^2 u_r}{\partial z^2} + \frac{\partial^2 u_r}{\partial r^2} + \frac{1}{r} \frac{\partial u_r}{\partial r} - \frac{u_r}{r^2}\right). \label{eq:momr}
\end{equation}
We introduce the velocities $U_z$ and $U_r$ and define the non-dimensional quantities as
\begin{equation*}
	\tilde{r} := \frac{r}{r_0}, \ \tilde{z} := \frac{z}{l_0}, \ \tilde{t} := t \frac{U_z}{l_0}, \ \tilde{u}_z := \frac{u_z}{U_z}, \ \tilde{u}_r := \frac{u_r}{U_r}, \ \tilde{P} := \frac{P}{\rho U_z^2}, \ \epsilon := \frac{U_r}{U_z}.
\end{equation*}
In a laminar flow the value of $U_r$ is often chose as the radial velocity of the arterial wall. We assume that these walls can only be subjected to small displacements, $U_r$ can be assumed to be small and therefore $\epsilon << 1$ \cite{womersley1957elastic}. Inserting the non-dimensional quantities into Equation\ref{eq:cont1} - Equation\ref{eq:momr} and dropping the terms of order $\epsilon^2$ or higher yields
\begin{equation}
	\frac{\partial (\tilde{r} \tilde{u}_r)}{\partial \tilde{r}} + \frac{\partial (\tilde{r} \tilde{u}_z)}{\partial\tilde{z}} = 0, \label{eq:cont2}
\end{equation}
\begin{equation}
	\tilde{r} \frac{\partial \tilde{u}_z}{\partial \tilde{t}} + \frac{ \partial (\tilde{r} \tilde{u}_z^2)}{\partial \tilde{z}} + \frac{ \partial (\tilde{r} \tilde{u}_z \tilde{u}_r)}{\partial \tilde{r}} + \tilde{r} \frac{\partial \tilde{P}}{\partial \tilde{z}} = \frac{\mu}{\rho} \frac{l_0}{U_z r_0^2} \frac{\partial}{\partial \tilde{r}} \left( \tilde{r} \frac{\partial \tilde{u}_z}{\partial \tilde{r}} \right), \label{eq:momz}
\end{equation}
\begin{equation*}
	\frac{\partial \tilde{P}}{\tilde{r}} = 0.
\end{equation*}
Introducing the non-dimensional vessel inner radius $\hat r$, the averaged velocity over the circular cross-section is defined:
\begin{equation*}
	\hat{u} := \frac{1}{\hat{r}^2} \int_0^{\hat{r}} 2 \tilde{u}_z \tilde{r} d\tilde{r},
\end{equation*}
and the Coriolis' coefficient:
\begin{equation*}
	\alpha := \frac{1}{\hat{r}^2 \hat{u}} \int_0^{\hat{r}} 2 \tilde{u}_z^2 \tilde{r} d\tilde{r}.
\end{equation*}
The Coriolis' coefficient parameter corrects for the fact that the momentum Equationno longer expresses the conservation of momentum but the conservation of the momentum averaged over the radial component\cite{article10002407}. Using these definitions and applying the no-slip boundary condition, while assuming negligible longitudinal wall displacement, we get
\begin{equation*}
	\tilde{u}_r|_{\tilde{r}=\hat{r}} = \frac{\partial \hat{r}}{\partial \tilde{t}},
\end{equation*}
and the equations \ref{eq:cont2} and \ref{eq:momz} become
\begin{equation}
	\hat{r} \frac{\partial \hat{r}}{\partial \tilde{t}} + \frac{1}{2}\frac{\partial}{\partial \tilde{z}} \left(  \hat{u}\hat{r}^2 \right) = 0, \label{eq:cont3}
\end{equation}
and
\begin{equation}
	\frac{(\partial \hat{r}^2 \hat{u})}{\partial \tilde{t}} + \frac{\partial \left( \alpha \hat{r}^2 \hat{u}^2 \right)}{\partial \tilde{z}} + \hat{r}^2 \frac{\partial \tilde{P}}{\partial \tilde{z}} = 2\frac{\mu}{\rho} \frac{l_0}{U_z r_0^2} \hat{r} \frac{\partial \tilde{u}_z}{\partial \tilde{z}} |_{\hat{r}},\label{eq:momz1}
\end{equation}
respectively.  Introducing a dimensional axial velocity in terms of the dimensional inner vessel radius $\bar{r}$
\begin{equation*}
	u := U_z \hat{u} = \frac{1}{\bar{r}^2} \int_0^{\bar{r}} 2su_zds,
\end{equation*}
allows us to write the Coriolis' coefficient in terms of dimensional variables:
\begin{equation*}
	\alpha = \frac{1}{\bar{r}^2 u^2} \int_0^{\bar{r}} 2ru_z^2dr.
\end{equation*}
In dimensional variables the continuity Equation(Equation\ref{eq:cont3}) and the $z$-momentum Equation(Equation\ref{eq:momz1}) are written 
\begin{equation*}
	\frac{\partial \bar{r}^2}{\partial t} + \frac{\partial (\bar{r}^2 u^2)}{\partial z} = 0,
\end{equation*}
and
\begin{equation*}
	\frac{\partial (\bar{r}^2 u)}{\partial t} + \frac{\partial (\alpha \bar{r}^2 u^2) }{\partial z} + \frac{\bar{r}^2}{\rho} \frac{\partial P}{\partial z} = 2 \frac{\mu}{\rho} \bar{r} \frac{\partial u_z}{\partial r} |_{\bar{r}},
\end{equation*}
respectively. Introducing the cross-sectional area $A := \pi \bar{r}^2$ and the volumetric flow-rate $Q := Au$ allows us to write the 1D model as:
\begin{equation}
		\frac{\partial A(z;t)}{\partial t} + \frac{\partial Q(z;t)}{\partial z} = 0, \label{eq:1deqs1}
\end{equation}
and
\begin{equation}
		\frac{\partial Q(z;t)}{\partial t} + \frac{\partial}{\partial z}\left(\alpha(z;t) \frac{Q(z;t)^2}{A(z;t)} \right) + \frac{A(z;t)}{\rho(z;t)} \frac{\partial P(z;t)}{\partial z} = 
		-2 \frac{\mu}{\rho(z;t)} ( \gamma + 2 ) \frac{Q(z;t)}{A(z;t)}. \label{eq:1deqs2}
\end{equation}
We have reintroduced the time and space dependency 
\begin{equation*}
		z \in [0,l],\  t \in \mathbb{R}_{\geq 0}, 
\end{equation*}
for clarity. Here $\gamma$ is the velocity profile parameter that determines the velocity profile by a common parametric approximation 
\begin{equation*}
	u_z(z,r;t) := \frac{\gamma + 2}{\gamma} u(z;t) \left[ 1 - \left( \frac{r}{\bar{r}(z;t)} \right) \right].
\end{equation*}
For this velocity profile the Coriolis coefficient is
\begin{equation*}
	\alpha = \frac{\gamma + 2}{\gamma + 1}.
\end{equation*}
Setting $\gamma=9$, $\alpha = \frac{11}{10}$ corresponds to a plug-like flow, a rather flat velocity profile that decays rapidly at the artery walls. Setting $\gamma=2$, $\alpha = \frac{4}{3} $ results in a parabolic velocity profile and hence a Poiseuille-like flow. \cite{köppl2023dimension} \cite{barnard1966theory}

We want to determine the variables $Q, A,$ and $P$. Until now we've considered the equations of mass and momentum conservation that allow computing $Q$ and $A$. To close the system and determine $P$, as well, another Equationis required. There are many different ways to model the pressure $P$ within a deforming tube, but we consider to model it using the purely elastic relation derived from the tube law \cite{alastruey2008reduced}. For a full comparison of a wide range of such models, we refer to \cite{gomez2017analysis}. The pressure $P$ is modeled through the external pressure $P_{ext}$ stemming from surrounding tissue, as well as, the pressure from the deformation of the vessel
\begin{equation}
	P_{def}(z;t) := \frac{h_0(z) \sigma(z;t)}{\pi r(z;t)}, \label{eq:pdef}
\end{equation}
given by the Young-Laplace law \cite{comte1799traite, thomas1805essay}.
In Equation\ref{eq:pdef},  $\sigma$ denotes the stress resulting from the internal forces that neighboring particles of the continuous material exert on each other. Assuming linear elasticity and using Hooke's law \cite{hooke1678lectures} we arrive at
\begin{equation}
	\sigma(z;t) := \epsilon(z;t) E \label{eq:stress},
\end{equation}
where $E$ the Young's modulus. The quantity $\epsilon$ is referred to as the stress and models the deformation of the material. Assuming an isotropic, homogeneous, and incompressible arterial wall, an axisymmetric deformation, and circular cross-sections that are independent of each other $\epsilon$ is given by
\begin{equation}
	\epsilon(z;t) := \frac{r(z;t)-r_0(z;t)}{ (1-\nu^2) r_0(z;t)}. \label{eq:strain}
\end{equation}
In this definition, $\nu$ is the elasticity parameter or Poisson's ratio. It is the ratio of transverse contraction to longitudinal extension strain in the direction of the stretching force.
Combining Equations \ref{eq:pdef}-\ref{eq:strain} with $P_{ext}$ we get \cite{sherwin2003one, sherwin2003computational}
\begin{align}
	P(z;t) &:= P_{ext}(z;t) + \beta \left( \sqrt{\frac{A(z;t)}{A_0(z)}}-1 \right),  \label{eq:p_tot}\\
	\beta(z) &:=  \frac{\sqrt{\pi} E h_0(z)}{(1-\nu^2) \sqrt{A_0(z)}}.
\end{align}
We will refer to $\beta$ as the elasticity coefficient from here on. Comparisons to experimental data have shown that a plug-like velocity profile provides a reasonable approximation of the velocity waveform compared to patient data  \cite{hunter1972numerical, smith2000generation, smith2002anatomically}.
Therefore, we set $\gamma = 9, \alpha = \frac{11}{10}$.
Since the plug-like flow velocity profile is relatively flat, the average of $Q$ doesn't differ from the actual value anywhere along $r$. Therefore, we can also set the Coriolis' coefficient $\alpha$ (averaging correction parameter) to $\alpha = 1$ \cite{formaggia2010cardiovascular}. Inserting these fixed values for $\gamma$ and $\alpha$ into Equations \ref{eq:1deqs1} and \ref{eq:1deqs2} leads to 
\begin{equation}
	\begin{aligned} 
		\frac{\partial A}{\partial t} + \frac{\partial Q}{\partial z} &= 0, \\ 
		\frac{\partial Q}{\partial t} + \frac{\partial}{\partial z}\left(\frac{Q^2}{A} \right) + \frac{A}{\rho} \frac{\partial P}{\partial z} &= -22 \frac{\mu}{\rho} \frac{Q}{A}.
	\end{aligned} \label{1deqs2}
\end{equation}
We drop the space and time dependence for simplifying the notation. We can rewrite Equation\ref{1deqs2} in it's conservative form by defining the following shorthands :
\begin{equation*}
	\mathbf{U} := 
	\begin{bmatrix}
		A \\
		Q
	\end{bmatrix},
\end{equation*}
\begin{equation*}
	\mathbf{F} \left( \mathbf{U} \right) := 
	\begin{bmatrix}
		Q \\
		\frac{Q^2}{A} + \frac{\beta A^{\frac{3}{2}}}{3\rho\sqrt{A_0}}
	\end{bmatrix},
\end{equation*}
\begin{equation*}
	\mathbf{S} \left( \mathbf{U} \right) := 
	\begin{bmatrix}
		0 \\
	-22\frac{\mu}{\rho}\frac{Q}{A} 
	\end{bmatrix},
\end{equation*}
where we have substituted $P$ with Equation\ref{eq:p_tot} and set $\nu = \frac{1}{2}$ since biological tissue is nearly incompressible \cite{sherwin2003one}. The conservative form then reads:
\begin{equation}
	\begin{aligned}
		\frac{\partial \mathbf{U}}{\partial t} + \frac{\partial \mathbf{F} \left( \mathbf{U} \right)}{\partial z} &= \mathbf{S} \left( \mathbf{U} \right), \ t>0, \ z \in \left[ 0,l \right], \\
		\mathbf{U} \left( z;0 \right) &= \mathbf{U}_0 \left( z \right), \ z \in \left[ 0,l \right], \\
		\mathbf{U} \left( 0;t \right) &= \mathbf{U}_L \left( t \right), \ t>0,\\
		\mathbf{U} \left( l;t \right) &= \mathbf{U}_R \left( t \right), \ t>0,
	\end{aligned} \label{eq:1deqs3}
\end{equation}
where we have added initial values $\mathbf{U}_0$, boundary conditions $\mathbf{U}_L$ and $\mathbf{U}_R$ so that Equation\ref{eq:1deqs3} now yields a well posed problem \cite{formaggia2010cardiovascular}. This concludes the derivation of the 1D-model for a single artery. The initial and boundary conditions in Equation\ref{eq:1deqs3} need to be modeled explicitly as well. We will address after introducing the considered numerical methods, and notation that will make it easier to describe the initial and boundary conditions.

\subsection{Choice of numerical scheme} \label{sec:lr}

The numerical solution of Equation\ref{eq:1deqs3} has been approached using four different strategies, namely the method of characteristics, the finite difference (FD) \cite{smith2002anatomically,elad1991numerical,elad1991numerical,pontrelli2003numerical,reymond2009validation}, finite volume (FV), or finite element (FE) methods. Using the method of characteristics, the hyperbolic PDE that describes the hemodynamics is rewritten as an ODE of the Riemann invariants of the system along the corresponding characteristic lines \cite{streeter1963pulsatile,bodley1971non,parker1990forward,wang2004wave,wang2003time}. The finite volume and the finite element methods have become more prevalent in recent years. The FV method is very suitable for modeling functions that have discontinuous behavior \cite{toro2001shock}, which means they can deal with potentially large gradients that appear in the solution \cite{shu1988efficient,harten1997uniformly}. In the presented formulation, vascular networks are modeled as individual vessels with discontinuous boundaries, and therefore an FV method would be beneficial. In the literature, both the finite volume \cite{melis2017gaussian,brook1999numerical,brook2002model} and the finite element methods \cite{sherwin2003one,sherwin2003computational,formaggia2001coupling,wan2002one,porenta1986finite,rooz1982finite,bessems2008experimental} have been considered.

There is also increasing complexity involved when implementing models that are based on the method of characteristics all the way up to finite element methods. For this work, we chose to implement the hemodynamics solver using an FV scheme, specifically using the MUSCL solver, because FV solvers offer many benefits when applied to hyperbolic systems. Equations \ref{eq:1deqs3} comprise a system of hyperbolic PDEs, which means that it describes phenomena where information travels at finite speeds.  The benefits relate to the shock capture and the inherently conservative properties of the method. They offer enough accuracy to model the problem at hand while having the added benefit of reduced implementation complexity when comparing them to an FE implementation. The numerical realization of the 1D model within a single vessel described in Section \ref{sec:sv} is achieved using a Finite Volume (FV) method to solve the homogeneous part of Equation\ref{eq:1deqs3}, while the source term is accounted for via a forward Euler time step. The FV method scheme considered in this work is the monotonic upstream-centered scheme for conservation laws (MUSCL). The initial data, inlets, junctions, and outlets of the system are set from data, in accordance with neighboring vessels, or through a reflective or a Windkessel model. 


\subsection{Finite Volume Method} \label{sec:fv}
To describe the FV method, we will first introduce the time-space discretization nomenclature. The space discretization is defined as follows
\begin{align*}
	&0 = z_0 < z_1 < z_2 < ... < z_N = l, \\
	&z_i :=  i \Delta z,\ i \in \{0,1,...,N\} , \\
	&\Delta z := \frac{l}{N}.
\end{align*}
A time-step series is defined as
\begin{align*}
	0 < t^1 < t^2 < t^3 < ...\ ,
\end{align*}
where the explicit computation of the values of $t^n$, $n \in \mathbb{N}$  will be given in \ref{sec:cfl}.
The balance Equationon the infinitesimal control volume defined as
\begin{align*}
	&[\tilde{z}, \tilde{z} + \delta z] \times [\tilde{t}, \tilde{t} + \delta t], \\
	&\tilde{z} \in (0,l), \\
	&\tilde{t} > 0,
\end{align*}
for a hyperbolic law in its conservation form \cite{guinot2012wave} (e.g. Equation\ref{eq:1deqs3}) is given by
\begin{equation*}
	\int_{\tilde{z}}^{\tilde{z} + \delta z} \mathbf{U} (z; \tilde{t} + \delta t) - \mathbf{U} (z;\tilde{t}) dz = \int_{\tilde{t}}^{\tilde{t}+\delta t} \mathbf{F}(\tilde{z};t) - \mathbf{F}(\tilde{z} + \delta z; t) dt   + \int_{\tilde{t}}^{\tilde{t} + \delta t} \int_{\tilde{z}}^{\tilde{z} + \delta z}  \mathbf{S}(z;t) dzdt.
\end{equation*}
We define cell $i$ as the control volume: 
\begin{align*}
	&[z_{i-\frac{1}{2}},z_{i+\frac{1}{2}}] \times [t^n, t^{n+1}], \\
	&\Delta t := t^{n+1}-t^n,
\end{align*}
where $z_{i \pm \frac{1}{2}}$ are the spatial points exactly in-between $z_{i}$, $z_{i+1}$ and $z_{i-1}$, $z_{i}$ respectively. The balance Equationis then written:
\begin{equation*}
	(\mathbf{U} _i^{n+1}-\mathbf{U} _i^n) \Delta z = (\mathbf{F}_{i-\frac{1}{2}}^{n+\frac{1}{2}} - \mathbf{F}_{i+\frac{1}{2}}^{n+\frac{1}{2}}) \Delta t + \mathbf{S}_i^{n+\frac{1}{2}} \Delta z \Delta t.
\end{equation*}
Here $\mathbf{U}_i^n$, $i \in \{0,1,...,N\}$ is the average of $\mathbf{U}$ over the spatial component of the cell $i$ at time-step $n$,
$\mathbf{F}_{i \pm \frac{1}{2}}^{n + \frac{1}{2}}$ is the average of the flux $\mathbf{F}$ where the cells $i$ and $i+1$, or  $i-1$ and $i$  meet exactly in-between the time-steps $n$ and $n + 1$, and $\mathbf{S}_i^{n+\frac{1}{2}}$ is the average of the source term $\mathbf{S}$ over the spatial component of the cell, again at the time-step right in-between $n$ and $n+1$. Dividing by $\Delta z$ and rearranging the terms provides:
\begin{equation}
	\mathbf{U} _i^{n+1} = \mathbf{U}_i^n + \frac{\Delta t}{\Delta z} (\mathbf{F}_{i-\frac{1}{2}}^{n+\frac{1}{2}} - \mathbf{F}_{i+\frac{1}{2}}^{n+\frac{1}{2}}) + \mathbf{S}_i^{n+\frac{1}{2}} \Delta t. \label{eq:fv}
\end{equation}
Therefore the value of $\mathbf{U}$ at time-step $n+1$ is determined from the value at the current time-step $n$ and the values of the flux $\mathbf{F}$ and the source term $\mathbf{S}$ at the intermediate time-step $n + \frac{1}{2}$ \cite{guinot2012wave}.

\subsection{The MUSCL Solver} \label{sec:muscl}
The MUSCL solver \cite{van1977towards, van1979towards} is commonly composed of 4 steps:
\begin{enumerate}
	\item Reconstructing a plausible $\mathbf{\hat{U}}(z;t^n)$ for $z \in [z_{i-\frac{1}{2}}, z_{i+\frac{1}{2}}]$ from $\mathbf{U}_i^n$.
	\item Limiting the profile of said reconstruction, as to be "Total Variation Diminishing" (TVD), this is often also referred to as slope limiting.
	\item Solving a generalized Riemann problem.
	\item Using the solution of this Riemann problem to compute $\mathbf{U}_i^{n+1}$, also referred to as balancing over the cells.
\end{enumerate}
We describe each step in more detail:
\paragraph{Reconstruction} 
When reconstructing a solution within a cell $\mathbf{\hat{U}}(z;t^n), z \in [z_{i-\frac{1}{2}}, z_{i+\frac{1}{2}}]$ with respect to the computed average of that cell $\mathbf{U}_i^n$, conservation must be taken into account. This means that the following should hold:
\begin{equation}
	\frac{1}{\Delta z} \int_{z_{i-\frac{1}{2}}}^{z_{i+\frac{1}{2}}} \mathbf{\hat{U}}_i^n(z;t^n) dz = \mathbf{U}_i^n,
\end{equation}
The reconstruction in the considered MUSCL scheme is chosen as the linear expression:
\begin{align*}
	\mathbf{\hat{U}}(z;t^n) &= \mathbf{U}_i^n + (z-z_i)a_i^n, \\
	a_i^n &= \frac{\mathbf{U}_{i+1}^n - \mathbf{U}_{i-1}}{z_{i+1} - z_{i-1}}.
\end{align*}
\paragraph{Profile Limiting} 
We define the total variation of the numerical solution $\mathbf{U}_i^n$ at time-step $n$ as 
\begin{equation*}
	TV(\mathbf{U})^n = \sum_{i=0}^{N-1} | \mathbf{U}_{i+1}^n - \mathbf{U}_{i}^n|.
\end{equation*}
A numerical scheme is "Total Variation Diminishing" (TVD) if the following holds
\begin{equation*}
	TV(\mathbf{U})^{n+1} \leq TV(\mathbf{U})^n.
\end{equation*}
According to \cite{van1977towards} and \cite{colella1984piecewise} the MUSCL scheme is TVD if it satisfies
\begin{align}
	\min (\mathbf{U}_{i-1}^n, \mathbf{U}_i^n) &\leq \mathbf{\hat{U}}_i^n(z_{i-\frac{1}{2}}) \leq \max (\mathbf{U}_{i-1}^n, \mathbf{U}_i^n), \label{eq:tvd1}\\
	\min (\mathbf{U}_{i}^n, \mathbf{U}_{i+1}^n) &\leq \mathbf{\hat{U}}_i^n(z_{i+\frac{1}{2}}) \leq \max (\mathbf{U}_{i}^n, \mathbf{U}_{i+1}^n).\label{eq:tvd2}
\end{align}
This can be achieved empirically via performing the following checks:
\begin{enumerate}
	\item The cell $i$ is not a local extremum, that is, $(\mathbf{U}_i^n - \mathbf{U}_{i-1}^n)(\mathbf{U}_{i+1}^n-\mathbf{U}_i^n) > 0$. \label{chk1}
	\item Equation\ref{eq:tvd1} is satisfied. \label{chk2}
	\item Equation\ref{eq:tvd2} is satisfied. \label{chk3}
\end{enumerate}
If Check \ref{chk1} is not satisfied, $a_i^n$ is set to zero. Then, Check \ref{chk2} and Check \ref{chk3} do not need to be checked. If  Check \ref{chk1} is satisfied but Check \ref{chk2} is not, $a_i^n$ is set to the highest value allowed by Equation\ref{eq:tvd1}
\begin{equation*}
	a_i^n = 2 \frac{\mathbf{U}_i^n - \mathbf{U}_{i-1}^n}{\Delta z}.
\end{equation*}
Otherwise, if Check \ref{chk3} is not satisfied, $a_i^n$ is set to the highest value allowed by Equation\ref{eq:tvd2}
\begin{equation*}
	a_i^n = 2 \frac{\mathbf{U}_{i+1}^n - \mathbf{U}_{i}^n}{\Delta z}.
\end{equation*}

\paragraph{Generalized Riemann Problem and Balancing} \label{ssec:grp}
To get an approximation of the flux, the Riemann problem
\begin{align*}
\frac{\partial \mathbf{U}}{\partial t} + \frac{\partial \mathbf{F}}{\partial z} &= 0, \\
\mathbf{U}(z,t^n) &= \left\{
\begin{array}{c}
    \mathbf{\hat{U}}_i^n(z_{i+\frac{1}{2}}) \text{, for } z < z_{i+\frac{1}{2}}, \\
    \mathbf{\hat{U}}_{i+1}^n(z_{i+\frac{1}{2}}) \text{, for } z > z_{i+\frac{1}{2}},
\end{array} \right.  
\end{align*}
has to be solved. This results in a flux $\mathbf{F^*}$ used to compute the averaged quantities at the next half time-step by Equation\ref{eq:fv}:
\begin{align*}
\mathbf{U}_{i+\frac{1}{2},L}^{n+\frac{1}{2}} &= \mathbf{\hat{U}}_i^n(z_{i+\frac{1}{2}}) + \frac{\Delta t}{2 \Delta z} \left( \mathbf{F^*}(\mathbf{\hat{U}}_i^n( z_{i-\frac{1}{2}})) - \mathbf{F^*}(\mathbf{\hat{U}}_i^n(z_{i+\frac{1}{2}}) \right),\\
\mathbf{U}_{i+\frac{1}{2},R}^{n+\frac{1}{2}} &= \mathbf{\hat{U}}_i^n(z_{i+\frac{1}{2}}) + \frac{\Delta t}{2 \Delta z} \left( \mathbf{F^*}(\mathbf{\hat{U}}_{i+1}^n( z_{i+\frac{1}{2}})) - \mathbf{F^*}(\mathbf{\hat{U}}_i^n(z_{i+\frac{3}{2}}) \right),
\end{align*}
where we have ignored the source term $\mathbf{S}$ for the moment. The resulting $\mathbf{U}_{i+\frac{1}{2},L}^{n+\frac{1}{2}}$ and $\mathbf{U}_{i+\frac{1}{2},R}^{n+\frac{1}{2}}$ are then used as left and right states of a new Riemann problem. The flux resulting from the Riemann problem provided in Equation\ref{eq:fv} provides the final value for $\mathbf{U}_i^{n+1}$. The fluxes in both Riemann problems are computed numerically using the conservative form of the Lax-Friedrichs scheme. Given that the flux function is Lipschitz continuous, this scheme is conservative and consistent \cite{köppl2023dimension}. Moreover, it is first-order accurate and stable for small enough step sizes (see \ref{sec:cfl}) \cite{leveque1992numerical}. We denote the approximation computed for $\mathbf{U}$ at time step $n+1$ by $\bar{\mathbf{U}}^{n+1}$.

\paragraph{Time-stepping the source term} \label{sec:fe}
To account for the source term, the following problem has to be solved
\begin{equation*}
\frac{d \mathbf{U}}{d t}=\mathbf{S}(\mathbf{U}) .
\end{equation*}
This is done here via the forward Euler method \cite{köppl2023dimension}
\begin{equation*}
\mathbf{U}^{n+1}=\bar{\mathbf{U}}^{n+1} + \Delta t \mathbf{S}\left(\bar{\mathbf{U}}^{n+1}\right).
\end{equation*}

\subsection{Initial and Boundary Conditions} \label{sec:icbc}
The left boundary condition of the inlet vessel is set by parameterizing the pulse wave flow from the heart. The boundary conditions for neighboring vessels are assumed based on conservation laws. The right boundary conditions for outlet vessels are set through either a reflection or a Windkessel model. The initial values of the quantities of interest are either assumed to be zero for the flow or a constant value measured from imaging, cross-sectional area.  

\subsubsection{Initial Conditions} \label{ssec:initial_conditions}
It is difficult to determine the initial conditions that are useful for a hemodynamics calculation. What is usually done therefore is to simply set
\begin{align*}
u(z;0) &\equiv 0, &z \in [0,l],\\
A(z;0) &= A_0(z), &z \in [0,l], \\
Q(z;0) &= u(z;0)A(z;0) \equiv 0, &z \in [0,l].
\end{align*}
The pressure $P$ is then calculated using Equation\ref{eq:p_tot}. These assumed initial conditions lead to incorrect values when propagated through the arterial network. For this reason, the output of the solver is fed back into the simulation as a new initial condition until a steady state has been reached. When the correct boundary values are established, two or three iterations of this process are needed to reach approximately a steady-state solution \cite{formaggia2010cardiovascular}. In the next subsection, we discuss how to set the boundary values. 

\subsubsection{Boundary Conditions} \label{ssec:boundary_values}

There are three possible types of boundary conditions depending on the position of the vessel in the arterial network, namely the left boundary condition for an inlet vessel, the right boundary condition of an outlet vessel, and boundary conditions for neighboring vessels, bifurcations or junctions. 
We discuss these types of boundary condition separately in the next paragraphs.

\paragraph{Inlet}\label{sssec:inlets}
The left boundary condition for the inlet vessel is usually computed by fitting a Fourier series, or similar, expansion to given data. This data represents either the pressure $P_L(t)$ or the volumetric flow $Q_L(t)$ coming from the heart and is considered in the system as follows:
\begin{align*}
P(0,t) &= P_L(t), &t \in [0,\infty), \\
Q(0,t) &= Q_L(t), &t \in [0,\infty).
\end{align*}
If one of these quantities is given at a time $t^{n+1}$ the remaining quantities describing our system can also be computed at $t^{n+1}$ using extrapolation of characteristics. Considering Riemann's method of characteristics \cite{riemann1860fortpflanzung} for the governing Equations \ref{eq:p_tot},\ref{eq:1deqs3} it is implied that changes in pressure and velocity propagate in the backward and forward direction at speeds $u-c$ and $u+c$. These quantities are commonly referred to as Riemann invariants due to being constant along the characteristic curves \cite{sarra2003method}. For a given state of our system $A$, $Q$, we denote the backward and forward Riemann invariants by
\begin{align*}
W_1(A,u) := u - c, \\
W_2(A,u) := u + c,
\end{align*}
where the wave speed $c(A) := \sqrt{\frac{\beta}{2\rho}\sqrt{A}}$. We define discretized Riemann invariants at time-step $n$ in cell $i$ along the lines of the discretization introduced in Section \ref{sec:fv} as
\begin{align*}
W_{1,i}^n := W_1(A_i^n,u_i^n) &= u^n_i - 4c^n_i,\\
W_{2,i}^n := W_2(A_i^n,u_i^n) &= u^n_i + 4c^n_i,
\end{align*}
where $c^n_i := \sqrt{\frac{\beta_i}{2\rho}\sqrt{A_i^n}}$. Therefore, at the first and the second cell the two Riemann invariants read as
\begin{align*}
W_{1,0}^n &= u^n_0 - 4c^n_0,\\
W_{2,0}^n &= u^n_0 + 4c^n_0,\\
W_{1,1}^n &= u^n_1 - 4c^n_1,\\
W_{2,1}^n &= u^n_1 + 4c^n_1.
\end{align*}
If the input data is given in as a volumetric flow, the Riemann invariants at the next time-step are computed via the following update formula
\begin{align}
W_{1,0}^{n+1} &= W^n_{1,0} + (W^n_{1,1} - W^n_{1,0})(c^n_0-u^n_0) \frac{\Delta t}{\Delta x} \label{update1a},\\
W_{2,0}^{n+1} &= \frac{2Q^{n+1}_0}{A^n_0} - W^{n+1}_{1,0}, \label{update2a}
\end{align}
where $Q^{n+1}_0$ is given through input data. This equates to linearly extrapolating the Riemann invariants at the first and second cell to meet the backwards and the forwards traveling characteristic curve of the second and first cell respectively. The missing input quantities are then computed from the new Riemann invariants as follows
\begin{align*}
u_0^{n+1} &= \frac{W_{1,0}^{n+1} + W_{2,0}^{n+1}}{2}, \\
c_0^{n+1} &= \frac{W_{2,0}^{n+1} - W_{1,0}^{n+1}}{4}, \\
A_0^{n+1} &= \frac{Q_0^{n+1}}{u_0^{n+1}}, \\
P_0^{n+1} &= P_{ext} + \beta \left( \sqrt{\frac{A_0^{n+1}}{A_{0,0}}} - 1 \right). 
\end{align*}
where $A_{0,0}$ is the reference cross-section at the cell $i=0$. If the input data is given as a pressure wave, the Riemann invariants at the next time-step are computed via the following update formula, again provided through a linear extrapolation
\begin{align}
W_{1,0}^{n+1} &= W^n_{1,0} + (W^n_{1,1} - W^n_{1,0})(c^n_0-u^n_0) \frac{\Delta t}{\Delta x} \label{update1b}, \\
W_{2,0}^{n+1} &= \frac{2Q^{n}_0}{A^n_0} - W^{n+1}_{1,0} \label{update2b}.
\end{align}
The missing input quantities are then computed from the new Riemann invariants as follows
\begin{align*}
u_0^{n+1} &= \frac{W_{1,0}^{n+1} + W_{2,0}^{n+1}}{2}, \\
c_0^{n+1} &= \frac{W_{2,0}^{n+1} - W_{1,0}^{n+1}}{4}, \\
A_0^{n+1} &= A_{0,0}\left(\frac{P_0^{n+1}-P_{ext}}{\beta}\right)^2,\\
Q_0^{n+1} &= A_0^{n+1} u_0^{n+1},
\end{align*}
where $P^{n+1}_0$ is as input data.

\paragraph{Reflection} \label{par:refl}
One of the most basic ways of setting outlet values is to consider coefficients to model the amount of wave that leaves the vessel and the amount that gets reflected. This is done equivalently to the inlet boundary conditions through linear extrapolation of characteristics
\begin{align}
W_{2,N}^{n+1}&=W_{2,N}^n+(W_{2,N-1}^n-W_{2,N-1}^n) (u_N^n+c_N^n) \frac{\Delta t}{\Delta x}, \label{eq:refl1}, \\
W_{1,N}^{n+1}&=W_{1,N}^0-R_t (W_{2,N}^{n+1}-W_{2,N}^0), \label{eq:refl2}
\end{align}
where $R_t \in [-1,1]$ is the reflection coefficient. At $R_t = 0$ no reflection occurs. It is noted that Equation\ref{eq:refl1} follows directly from linear extrapolation, while Equation\ref{eq:refl2} is constructed to transfer a part of the forward characteristic curve, backward as a reflection. The simulation quantities can be computed from the Riemann invariants as follows
\begin{align*}
u_N^{n+1} &= \frac{W_{1,N}^{n+1} + W_{2,N}^{n+1}}{2}, \\
c_N^{n+1} &= \frac{W_{2,N}^{n+1} - W_{1,N}^{n+1}}{4}, \\
Q_N^{n+1} &= A_N^n u_N^{n+1}.
\end{align*}
A more accurate method of computing the outlet boundary values is to use a Windkessel model.

\paragraph{Windkessel} \label{par:wk}
In cases where this method does not provide enough accuracy, Windkessel models are considered \cite {formaggia2010cardiovascular}. The three-element Windkessel model or RCR model is applied to simulate perfusion as described in \cite{CiCP-4-317}. Such models are referred to as lumped-parameter or 0D-models due to the lack of a spatial dependency. The concept behind this method of computing outlet values is the assumption of a state $A^*$, $u^*$ at the time step $n+1$ that results from the two states $A_L$, $u_L$ and $A_R$, $u_R$ at the time step $n$ to the left and right, respectively. The so-called virtual state $A^*$, $u^*$ therefore satisfies the following equations in its Riemann invariants
\begin{align}
W_1(A^*,u^*) = W_1(A_R,u_R), \label{eq:vrmi1}\\
W_2(A^*,u^*) = W_2(A_L,u_L). \label{eq:vrmi2}.
\end{align}
For a model that considers a terminal resistance, the virtual state has to satisfy
\begin{equation}
A^*u^* = \frac{P(A^*) - P_{ven}}{R}, \label{eq:tr1}
\end{equation}
where the constant $P_{ven}$ represents the venous pressure and $R$ models the resistive load resulting from omitted arteries, the capillary bed, and the venous network. The venous pressure in a healthy patient is only $8-12$ mmHg \cite{klingensmith2008washington}. In our simulation, it was therefore usually set to $0$. Inserting  Equations \ref{eq:p_tot} and \ref{eq:vrmi2} into Equation\ref{eq:tr1}:
\begin{equation}
f(A^*) := R \left(u_L + 4c(A_L) \right) A^* - 4Rc(A^*)A^* - \frac{\beta}{A_0} \left( \sqrt{A^*} - \sqrt{A} \right) P_{ven} = 0,\label{eq:tr2}
\end{equation}
where the wave speed $c$ of a cross-section $A$ is defined in Section \ref{sssec:inlets}. An additional term $P_{ext}$ for modeling exterior tissue pressure is added yielding
\begin{equation}
f(A^*) := R \left(u_L + 4c(A_L) \right) A^* - 4Rc(A^*)A^* - \frac{\beta}{A_0} \left( \sqrt{A^*} - \sqrt{A} \right) P_{ven} - P_{ext} = 0.\label{eq:tr3}
\end{equation}

The compliance of the omitted vessels and the resistance of the capillaries is modeled by a $CR_2$ system
\begin{equation*}
C \frac{dP_C}{dt} = A^*u^* - \frac{P_C-P_{ven}}{R_2},
\end{equation*}
where $C$ represents the compliance and $P_C$ is the pressure at said compliance. This Equationcan be discretized using an explicit Euler scheme to compute the values of $P_C$
\begin{equation}
P_C^n = P_C^{n-1} + \frac{\Delta t}{C} \left(A_Lu_L - \frac{ \left( P_C^{n-1} - P_{ven} \right)}{R_2} \right). \label{eq:pc}
\end{equation}
For the wave exiting the vessel to arrive at the compliance without being reflected, we couple the terminal resistance model with the $CR_2$ model. This is done by setting $R=R_1$ and $P_{ven}=P_C$ in Equation\ref{eq:tr2}. Since $A_L$, $u_L$ represents the state at the end of the vessel, they are set to $A_i^n$, $u_i^n$.  Equation\ref{eq:tr1} coupled with Equation\ref{eq:pc} is solved using a Newton solver \cite{atkinson1991introduction} to calculate $A^*$. Using Equations \ref{eq:p_tot}, \ref{eq:tr1} and the calculated value for $A^*$ we compute $u^*$. Finally, $A_i^{n+1}$, and $u_i^{n+1}$ are set to $A^*$, $u^*$, respectively \cite{alastruey2008reduced}. The parameters $R_1$, $C$, $R_2$ need to be determined for each individual patient. Different patients have potentially very different RCR parameters, and their accurate estimation is not trivial \cite{köppl2023dimension}. This is a prime example of justifying the need for efficient calibration algorithms such as the one provided in this manuscript. In the next section, we will go over how to set boundary values for junctions between vessels.

\subsection{Modeling Different Junction Types} \label{sssec:junctions}
Three types of junctions are considered in this manuscript. Conjunctions: joining two vessels; bifurcations: splitting one vessel into two; and anastomoses: joining two vessels into one. We will describe the governing equations of the junction types in the next three paragraphs. 

\paragraph{Conjunctions} \label{par:conjunctions}
A conjunction is the outlet of a vessel connected to the input of another vessel. For all types of junction, three types of Equationare taken into account: conservation of mass, pressure conservation, and extrapolation of characteristics. For the conjunctions, the conservation of mass leads to 
\begin{equation*}
A_{1,N}^{n+1} u_{1,N}^{n+1} - A_{2,0}^{n+1} u_{2,0}^{n+1}=0,
\end{equation*}
where $A_{1,N}^{n+1}$, $u_{1,N}^{n+1}$ and $A_{2,0}^{n+1}$, $u_{2,0}^{n+1}$ are the cross-section and the velocity at the outlet of the first vessel and the inlet of the second vessel respectively. The pressure conservation is given by 
\begin{equation*}
\beta_{1,N} \left( \frac{\sqrt{A_{1,N}^{n+1}}}{\sqrt{A_{0,1,N}}}-1\right)+\frac{1}{2} \rho \left( u_{1,N}^{n+1} \right) ^2 
-\beta_{2,0} \left( \frac{\sqrt{A_{2,0}^{n+1}}}{\sqrt{A_{0,2,0}}}-1\right)+\frac{1}{2} \rho \left( u_{2,0}^{n+1} \right)=0,
\end{equation*}
where $\beta_{1,N}$, $A_{0,1,N}$ and $\beta_{2,N}$, $A_{0,2,0}$ are the elasticity coefficient and the reference cross-section at the outlet of the first vessel and the inlet of the second vessel, respectively. The first and second terms represent the pressure coming from the arterial wall. The second and third terms describe the kinetic pressure that comes from the fluid itself. Finally, we introduce the virtual state $A^*$, $u^*$ generated from the two states $A_L$, $u_L$ and $A_R$, $u_R$
\begin{align*}
W_1^* := W_1(A^*,u^*) = W_1(A_R,u_R) = W_1(A_{2,0}^{n+1}, u_{2,0}^{n+1}), \\
W_2^* := W_2(A^*,u^*) = W_2(A_L,u_L) = W_1(A_{1,N}^{n+1}, u_{1,N}^{n+1}).
\end{align*}
This also equates to 
\begin{align*}
W_1^* = u_{2,0}^{n+1} - 4c_{2,0}^{n+1} = u_{2,0}^{n+1} - 4k_{2,0} \left( A_{2,0}^{n+1} \right)^{\frac{1}{4}}, \\
W_2^* = u_{1,N}^{n+1} + 4c_{1,N}^{n+1} = u_{1,N}^{n+1} + 4k_{1,N} \left( A_{1,N}^{n+1} \right)^{\frac{1}{4}},
\end{align*}
where $k_{l,i} := \sqrt{\frac{\beta_{l,i}}{2\rho}}$ for $l \in \{1,2\}$, $i \in \{0,1,...N\}$. This leads to the system of equations
\begin{equation*}
\mathbf{f}(\mathbf{q}) := \left[\begin{array}{c}
        q_{1}+4 k_{1,N} q_{3}-W_{2}^* \\
        q_{2}-4 k_{2,0} q_{4}-W_{1}^* \\
        q_{1} q_{3}^4-q_{2} q_{4}^4 \\
        \beta_1\left(\frac{q_{3}^2}{A_{0,1,N}^{1 / 2}}-1\right)+\frac{1}{2} \rho q_{1}^2-\beta_2\left(\frac{q_{4}^2}{A_{0,2,0}^{1 / 2}}-1\right)-\frac{1}{2} \rho q_{2}^2
\end{array}\right] \equiv 0. 
\end{equation*}
We simplify the notation in a matrix-vector form:
\begin{align*}
\mathbf{q}&:= \{q_k\}_{k=1}^4 = \left[\begin{array}{llll}
u_{1,N}^{n+1}, & u_{2,0}^{n+1}, & \left( A_{1,N}^{n+1}\right)^{1 / 4}, & \left( A_{2,0}^{n+1} \right)^{1 / 4}
\end{array}\right]^T. 
\end{align*}
The system is solved iteratively using the Newton method
\begin{align*}
\mathbf{J} \cdot \delta \mathbf{q}^m &= -\mathbf{f}\left(\mathbf{q}^m\right), \\
\mathbf{q}^{m+1} &= \mathbf{q}^m+\delta \mathbf{q}^m,
\end{align*}
where $\mathbf{J}$ is the Jacobian
\begin{equation*}
\mathbf{J}:=\left[\begin{array}{c c c c}{{{1}}}&{{0}}&{{{4}k_{1,N}}}&{{0}}\\ {{0}}&{{{1}}}&{{0}}&{{-4k_{2,0}}}\\ {{q_{c3}^{4}}}&{{-q_{c4}^{4}}}&{{4q_{c3}q_{3}^{3}}}&{{-4q_{c4}q_{c}^{3}}}\\ {{q_{c1}^{4}}}&{{-\rho q_{c2}}}&{{2\beta_{1}\frac{q_{c}}{A_{0,1,N}^{1/2}}}}&{{-2\beta_{2}\frac{q_{4}}{A_{0,2,N}^{42}}}}\end{array}\right].
\end{equation*}

The quantities $W_{1,N}^*$ and $W_{2,0}^*$ in $\mathbf{f}$ are calculated using the updated $\mathbf{q}$ at each iteration. The initial condition for the Newton solver is set as
\begin{align*}
\mathbf{q}^0&:=\left[\begin{array}{llll}
u_{1,N}^{n}, & u_{2,0}^{n}, & \left( A_{1,N}^{n}\right)^{1 / 4}, & \left( A_{2,0}^{n} \right)^{1 / 4}
\end{array}\right]^T.
\end{align*}
It is noted that the Jacobian only needs to be evaluated at $\mathbf{q}^0$ and can be reused at each iteration. This reduces computational cost and is commonly referred to as a frozen Newton method \cite{jin2010class,amat2013maximum,amat2018two}. 

\paragraph{Bifurcations} \label{par:bifurcations}
A bifurcation is a junction where one vessel outlet connects to two vessel inlets. For bifurcations, the same conservation laws are applied as for the conjunction, i.e. conservation of mass, conservation of pressure, and compatibility of the characteristic curves. This leads to the following system of equations
\begin{equation}
\mathbf{f}(\mathbf{q}):=\left[\begin{array}{c}
    q_{1}+4 k_{1,N} q_{4}-W_{1,N}^* \\
    q_{2}-4 k_{2,0} q_{5}-W_{2,0}^* \\
    q_{3}-4 k_{2,0} q_{6}-W_{3,0}^* \\
    q_{1} q_{4}^4-q_{2} q_{5}^4-q_{3} q_{6}^4=0 \\
    \beta_1\left(\frac{q_{4}^2}{A_{0,1,N}^{1 / 2}}-1\right)-\beta_2\left(\frac{q_{5}^2}{A_{0,2,0}^{1 / 2}}-1\right) \\
    \beta_1\left(\frac{q_{4}^2}{A_{0,1,N}^{1 / 2}}-1\right)-\beta_3\left(\frac{q_{6}^2}{A_{0,3,0}^{1 / 2}}-1\right) 
\end{array}\right] \equiv 0. \label{syseq_bif}
\end{equation}
The notation can be further simplified considering the shorthand:
\begin{equation*}
\mathbf{q}:=\{q_k\}_{k=1}^6 = \left[
\begin{array}{lll}
    u_{1,N}^{n+1}, & u_{2,0}^{n+1}, & u_{3,0}^{n+1},
\end{array} \right. 
\left. \begin{array} {lll}    \left(A_{1,N}^{n+1}\right)^{\frac{1}{4}}, & \left(A_{2,0}^{n+1}\right)^{\frac{1}{4}}, & \left(A_{3,0}^{n+1}\right)^{\frac{1}{4}}
\end{array}\right].
\end{equation*}
The Jacobian here reads as
\begin{equation*}
    \mathbf{J}:=\left[\begin{array}{c c c c c c}
            1&0&0&4k_{1,N}&0&0\\
            0&1&0&0&-4k_{2,0}&0\\
            0&0&1&0&0&-4k_{3,0}\\
            q_{4}^4 & -q_{5}^4 & -q_{6}^4 & 4q_{1}q_{4}^3 & -4q_{2}q_{5}^3 & -4q_{3}q_{6}^3 \\
            0 & 0 & 0 & 2\beta_1\frac{q_{4}}{A_{0,1,N}^{\frac{1}{2}}} & -2\beta_2\frac{q_{5}}{A_{0,2,0}^{\frac{1}{2}}} & 0\\
            0 & 0 & 0 & 2\beta_1\frac{q_{4}}{A_{0,1,N}^{\frac{1}{2}}} & -2\beta_3\frac{q_{6}}{A_{0,3,0}^{\frac{1}{2}}} & 0\\
        \end{array} 
    \right].
\end{equation*}
In analogy to the conjunction,  the frozen Newton method is applied to solve the system. In Equation\ref{syseq_bif} the kinetic energy term $\frac{1}{2} \rho u^2$ is neglected since their contribution to total pressure is small compared to the other terms \cite{Formaggia2003OnedimensionalMF}. 

\paragraph{Anastomosis} \label{par:anastomosis}
An anastomosis describes two outlet points of the vessel that connect to a single vessel input. For the anastomosis, applying the extrapolation of characteristics, conservation of mass, and pressure conservation leads to the following system of equations
\begin{equation*}
    \mathbf{f}(\mathbf{q}):=\left[\begin{array}{c}
            q_{1}+4 k_{1,N} q_{4}-W_{1,N}^* \\
            q_{2}+4 k_{2,N} q_{5}-W_{2,N}^* \\
            q_{3}-4 k_{2,0} q_{6}-W_{3,0}^* \\
            q_{1} q_{4}^4+q_{2} q_{5}^4-q_{3} q_{6}^4=0 \\
            \beta_1\left(\frac{q_{4}^2}{A_{0,1,N}^{1 / 2}}-1\right)-\beta_3\left(\frac{q_{6}^2}{A_{0,3,0}^{1 / 2}}-1\right) \\
            \beta_2\left(\frac{q_{5}^2}{A_{0,2,N}^{1 / 2}}-1\right)-\beta_3\left(\frac{q_{6}^2}{A_{0,3,0}^{1 / 2}}-1\right)
    \end{array}\right] \equiv 0.
\end{equation*}
Simplifying the notation again leads to
\begin{equation*}
    \mathbf{q}:=\{q_k\}_{k=1}^6=\left[
        \begin{array}{lll}
            u_{1,N}^{n+1}, & u_{2,N}^{n+1}, & u_{3,0}^{n+1}, 
    \end{array} \right. 
    \left. \begin{array}{lll}
            \left(A_{1,N}^{n+1}\right)^{\frac{1}{4}}, & \left(A_{2,N}^{n+1}\right)^{\frac{1}{4}}, & \left(A_{3,0}^{n+1}\right)^{\frac{1}{4}}
    \end{array} \right].
\end{equation*}
The Jacobian considered in the frozen Newton method reads:
\begin{equation*}
    \mathbf{J}:=\left[\begin{array}{c c c c c c}
            1&0&0&4k_{1,N}&0&0\\
            0&1&0&0&4k_{2,N}&0\\
            0&0&1&0&0&-4k_{3,0}\\
            q_{4}^4 & q_{5}^4 & -q_{6}^4 & 4q_{1}q_{4}^3 & 4q_{2}q_{5}^3 & -4q_{3}q_{6}^3 \\
            0 & 0 & 0 & 2\beta_1\frac{q_{4}}{A_{0,1,N}^{\frac{1}{2}}} & 0 & -2\beta_3\frac{q_{6}}{A_{0,3,0}^{\frac{1}{2}}} \\
            0 & 0 & 0 & 0 & 2\beta_2\frac{q_{5}}{A_{0,2,N}^{\frac{1}{2}}} &  -2\beta_3\frac{q_{6}}{A_{0,3,0}^{\frac{1}{2}}} \\
        \end{array} 
    \right].
\end{equation*}
As in the case of bifurcations, the kinetic energy terms in the pressure conservation equations are neglected. In the following section, we describe how to determine the time-step size to ensure the stability of the numerical scheme introduced in Sections \ref{sec:fv} and \ref{sec:muscl}.

\subsection{The CFL Condition} \label{sec:cfl}
The CFL condition is set, so the information traveling along a characteristic curve only affects neighboring cells. The reason for this is that the numerical scheme presented in Sections \ref{sec:fv} and \ref{sec:muscl}, only takes neighboring cells into account, thus considering information from further cells would result in the accumulation of numerical errors. The space discretization parameter $\Delta x$ is then set depending on how fine the spatial resolution is needed to have an accurate representation of the solution. Granted this, and for preventing the characteristic curves from reaching beyond the neighboring cell within a $\Delta t$, the following relation is considered
\begin{equation*}
    \Delta t \leq C_{CFL} \frac{\Delta x}{S_{max}},\hspace{10pt}  C_{CFL} \in (0,1),
\end{equation*}
where $S_{max} = \max_{x \in [0,L], v \in V} | u_v(x) + c_v(x) |$ and $V$ are the set of all vessels in the simulation.  In our case, $\Delta x$ was chosen to be at least $10^{-3}m$ and $C_{CFL}$ was set to $0.9$.

\section{Results} \label{sec:resl}

In this section, we present the results produced with our JAX implementation of the hemodynamics solver described in Section \ref{sec:1dm}, which we call jaxFlowSim. First, we describe how we convert the data from 3D geometries provided by the Vascular Model Repository (VMR) \cite{vascularmodel} into configurations for our solver. Then, we validate the proposed solver against an open-source implementation \cite{openBF} and compare the wall-clock time between the proposed and the validation implementations. Subsequently, in Section \ref{sec:da} we present results on different anatomies using the proposed methodology. In Section \ref{sec:sc}, we discuss scaling to a higher number of vessels with respect to the compile time and run time of our code. Finally, in Section \ref{sec:pi}, we describe different strategies for using the differentiability of the solver to perform parameter inference.

\subsection{Considering Different Anatomies from the Vascular Model Repository} \label{sec:ing}
In order to have a larger variety of vessel networks to test the proposed methodology, we developed an end-to-end process that accepts 3D VMR geometries and generates configuration files for our solver. In a first step, the 3D geometries are downloaded from the VMR and loaded into the 3D Slicer software. Here, the Vascular Modeling Tool Kit (VMTK) plugin is used to extract a 1D geometry in the form of a table.
 Using a python script, the vessel wall thickness is computed from the vessel radius using the empirical expression
\begin{equation*}
	h_0 = r_0 \left(a \exp(b r_0) + c \exp(d r_0)\right),
\end{equation*}
taken from Blanco et. al. \cite{blanco2014anatomically} with parameters
\begin{equation*}
	a = 0.2802, \ b = -5.053 \text{cm}^{-1}, \ c = 0.1324,\  d = -0.1114 \text{cm}^{-1}.
\end{equation*}
The Young's modulus can then be estimated from the vessel radius and the vessel wall thickness as follows
\begin{equation*}
	E = \frac{r_0}{h_0} k_1 \exp(k_2 r_0) + k_3,
\end{equation*}
given in Ottensen et. al. \cite{ottesen2004applied} with values
\begin{align*}
	k_1 &= 2.00 \times 10^7 \frac{\text{g}}{\text{s}^2\text{cm}}, \\
	k_2 &= -22.53 \frac{1}{\ \text{cm}}, \\
	k_3 &= 8.65 \times 10^5 \frac{\text{g}}{\text{s}^2\text{cm}}.
\end{align*}
By comparing the start (StartPointPosition) and end (EndPointPosition) points provided in the table generated by the VMTK plugin, our Python script also computes the connectivity of our vessel network. For the inlet of our network we can refer back to the data provided directly from the VMR which includes this quantity. At the outlets, the outlet parameters are set through a common approximation. In this approximation the total resistance of the system is defined as:
\begin{equation*}
	R_{tot} := \frac{P_{mean}}{Q_{car}},
\end{equation*}
where the mean pressure $P_{mean}$ and the cardiac output $Q_{car}$ are set to the nominal values:
\begin{align*}
	P_{mean} &:= \frac{120 \text{mmHg}}{80 \frac{\text{ml}}{\text{s}}} \approx 2 \times 10^8 \frac{\text{kg}}{\text{m}^4\text{s}}, \\
	Q_{car} &:= 80 \frac{\text{ml}}{\text{s}} = 8 \times 10^{-5} \frac{\text{m}^3}{\text{s}}. \\
\end{align*}
The total compliance is set as
\begin{align*}
	C_{tot} &:= 10^{-3} \frac{\text{cm}^5}{\text{dyne}} = 10^{-8} \frac{\text{m}^3}{\text{Pa}}.
\end{align*}
The individual resistances and the compliances are then set employing the rules for a parallel circuit. Let $I$ be an index set containing all vessels with a Windkessel outlet and $A_{o,i}, i \in I$ the reference cross section at the outlet of the $i$-th vessel. The resistance and the compliance at vessel $i$ are then computed as
\begin{align*}
	\xi_i &:= \frac{\sum_{j \in I} A_{0,j}}{A_{0,i}}, \\
	R_{tot,i} &:= \xi_i R_{tot}, \\
	C_{tot,i} &:= \frac{1}{\xi_i} C_{tot}. 
\end{align*}
Finally, the proximal and distal resistances $R_{1,i}, R_{2,i}$ need to be set for each vessel $i$. This is done by invoking an approximate ratio of about 1:10 between the two
\begin{align*}
	R_{1,i} &:= \frac{9}{100} R_{tot,i}, \\
	R_{2,i} &:= \frac{91}{100} R_{tot,i}.
\end{align*}
as explained in the SimVascular documentation \cite{simvascular}. Global simulation parameters such as blood viscosity and blood density can be read from the VMR provided data as well to complete the configuration that can now be provided to our 1D-hemodynamics solver.

\subsection{Comparison to the openBF solver} \label{sec:val}
\begin{figure} [h!]
\centering
\begin{subfigure}{0.4\textwidth}
  \includegraphics[width=\linewidth]{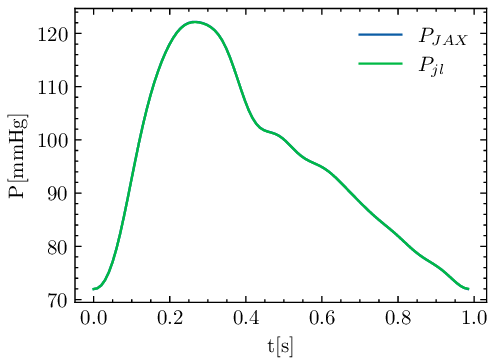}  
  \caption{Right Subclavian, Aortic Arch geometry.}
\end{subfigure}
\begin{subfigure}{0.4\textwidth}
  \includegraphics[width=\linewidth]{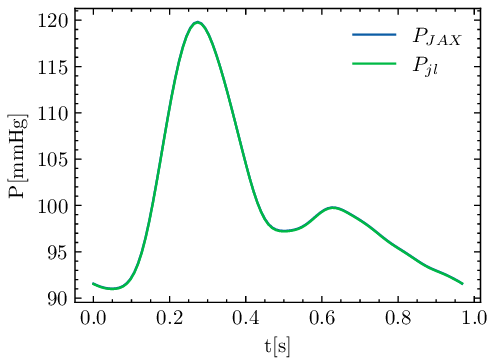}  
  \caption{Celiac Branch, Abdominal Aorta geometry.}
\end{subfigure}
\begin{subfigure}{0.4\textwidth}
  \includegraphics[width=\linewidth]{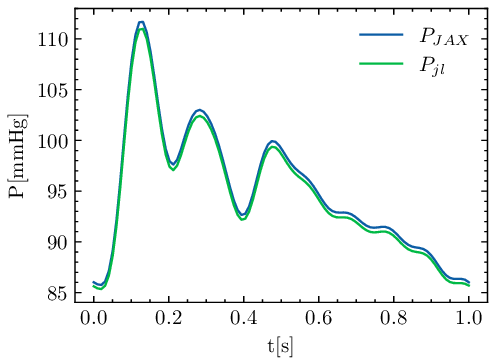}  
  \caption{Basilar, Cerebral geometry.}
\end{subfigure}
\begin{subfigure}{0.4\textwidth}
  \includegraphics[width=\linewidth]{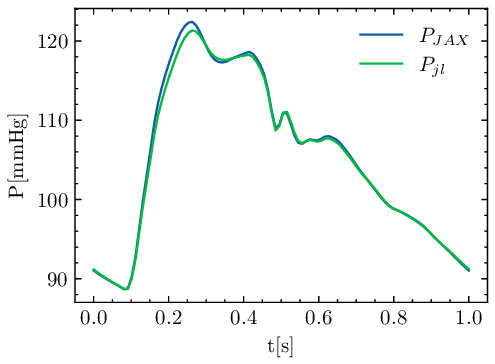}  
  \caption{Common Hepatic, ADAN56 geometry.}
\end{subfigure}
\caption{We provide comparisons between the pressure obtained from the openBF and the proposed solver for arteries from different anatomies.}
\label{fig:val}
\end{figure}

To validate the proposed solver, we compared it to the openBF 1D hemodynamic solver \cite{openBF} written and validated by Melis et al. \cite{melis2017gaussian}. That code is written in the Julia programming language \cite{julia}. We chose four vascular networks consisting of 9, 17, 19, and 77 vessels each for validation and compared the pressure over one cardiac cycle for a randomly selected vessel. The anatomies considered are an Aortic (AO), an Abdominal Aortic (ABAO), a Cerebral (CERE), and the ADAN56 full-body hemodynamic model \cite{melis2017gaussian}.  AO, ABAO, and CERE anatomies are downloaded from the VMR, an open online library consisting of three-dimensional anatomies of healthy patients and patients with different disorders, such as aneurysms or stenosis. The anatomies in the VMR library are named in the format (Patient Indicator)\_(Species)\_(Anatomy Indicator)\_(Disease Indicator). For example, 0007\_H\_AO\_H refers to the anatomy of the aorta of a patient with indicator 0007, who is a healthy human. In order to use the anatomies provided by the VMR, the method described in \ref{sec:ing} is used to generate the geometry needed by the 1D Navier-Stokes model. The left boundary conditions are taken directly from the VMR library for each anatomy. For ADAN56 anatomy, the left boundary condition for this case is found in Murgo et al. \cite{murgo1980aortic}.
The ADAN56 model, or the "anatomically detailed arterial network", is made up of the largest arteries of the human circulation. The model was developed by Blanco et al. \cite{blanco2014anatomically,blanco2014blood} and this specific variant of it contains 56 vessels. The 56 vessels are divided into 61 arterial segments which in turn are split into a total of 77 segments for simulation purposes. The comparison between the two solvers is presented in Figure \ref{fig:val}, where $\text{P}_\text{JAX}$ refers to the pressure computed by the proposed solver and $\text{P}_\text{jl}$ refers to the one computed using openBF.

\begin{figure}[h!]
	\centering
	\includegraphics[width=0.5\columnwidth]{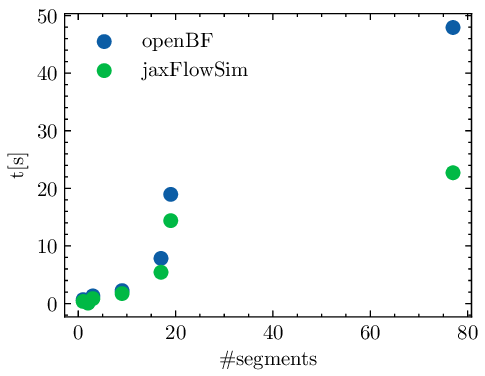}
	\caption{We present a comparison between the execution wall-clock time of openBF and jaxFlowSim in seconds.}
	\label{fig:comparison}
\end{figure}

The plots show that for each model there is barely any difference in the output of the two vascular simulations. Furthermore, there is a smaller than $1\%$ relative error difference between openBF and the proposed implementation, which validates our implementation; see Table \ref{tab:errors_comp}. We compare the wall-clock time for each of the solvers considering varying numbers of vessel segments. For a fair comparison, each solver is used 10 times for each anatomy, and the computation time is calculated as the average across these cases. The results are presented in \ref{fig:comparison}. We observe that for smaller numbers of vessel segments $(<10)$ the two implementations perform similarly, but for a larger number of vessel segments $(>15)$ the proposed implementation is more efficient. For the largest anatomy considered, the proposed implementation is almost twice as fast. These comparisons were performed on a personal computer with an AMD Ryzen 9 5950X CPU and 16 GB of RAM.

\begin{table}[h!]
    \centering
    \begin{tabular}{cccc}
    \hline
         & $\#$ Vessels & Artery Name & Error  ($\times 10^{-3}$)\\
    \hline
        AO & 9 & Right Subclavian & $0.25$\\
    \hline
        ABAO & 17 & Celiac Branch &$0.51$ \\
    \hline
        CERE & 19 & Basilar &$5.4$ \\
    \hline
        ADAN56 & 77 & Common Hepatic &$5.5$ \\
    \hline
    \end{tabular}
    \caption{We present comparisons using the relative $L_1$ error for the anatomies and the arteries presented in Figure \ref{fig:val}.}
    \label{tab:errors_comp}
\end{table}

\subsection{Results for Different Anatomies} \label{sec:da}
\begin{figure} [H]
	\centering
	\includegraphics[width=0.8\columnwidth]{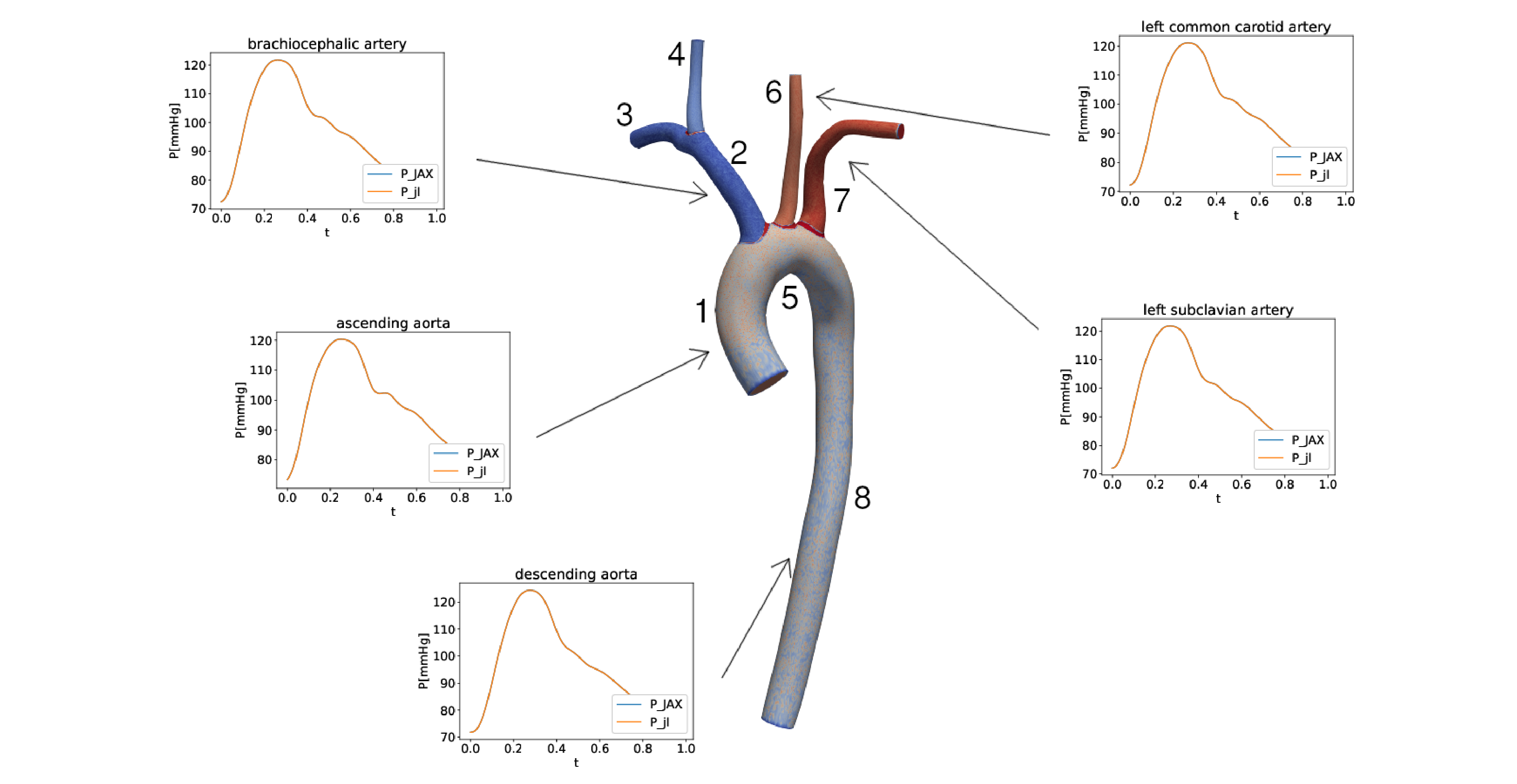}
	\caption{We present the jaxFlowSim results for different arteries of the Aorta anatomy.}
	\label{fig:aorta}
\end{figure}

In this section, we present the pressure waves provided by our implementation on the four different anatomies discussed before. The results of the first model of the VMR library, the aorta model, can be seen in Figure \ref{fig:aorta}. The model represents the ascending, the descending, the aortic arch, as well as the brachiocephalic, the right and left subclavian, and the common carotid arteries, see Table \ref{tab:aort}.
\begin{table}[h!] 
	\begin{center}
		\begin{tabular}{|c|l|c|l|} 
			\hline
			Legend & Vessel Name &Legend & Vessel Name\\
			\hline
			$\mathbf{1}$& ascending aorta &
			$\mathbf{5}$& arch of aorta \\
			$\mathbf{2}$& brachiocephalic artery &
			$\mathbf{6}$& left common carotid artery \\
			$\mathbf{3}$& right subclavian artery &
			$\mathbf{7}$& descending aorta \\
			$\mathbf{4}$& right common carotid artery &
			$\mathbf{8}$& left subclavian artery \\
			\hline
		\end{tabular}
	\end{center}
	\caption{This table lists the 8 arteries of the aorta model displayed in Figure \ref{fig:aorta}.}
	\label{tab:aort}
\end{table}

\begin{figure} [h!]
	\centering
	\includegraphics[width=0.8\columnwidth]{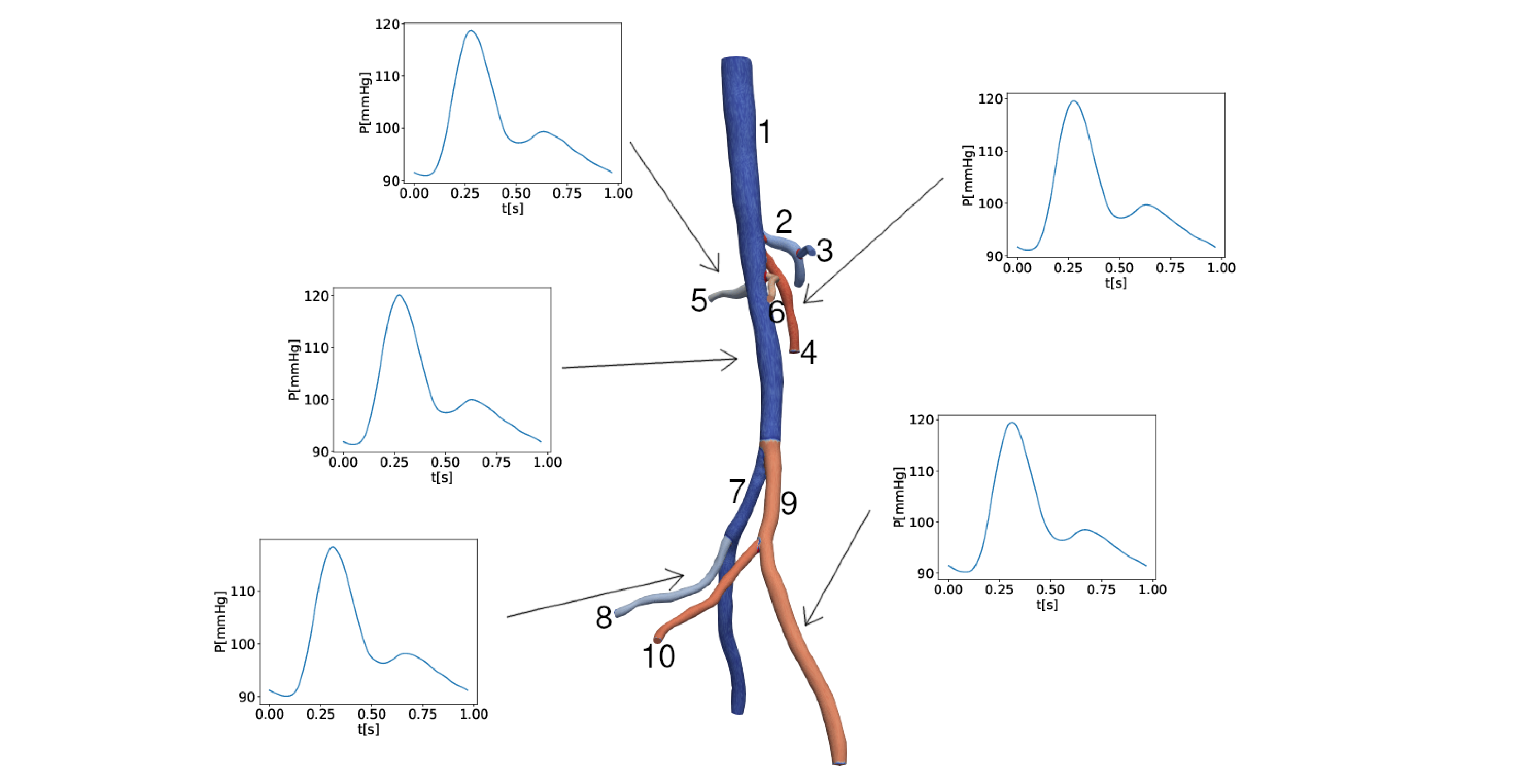}
	\caption{We present the jaxFlowSim results for different arteries of the Abdominal Aorta anatomy.}
	\label{fig:abdominal}
\end{figure}

\begin{figure} [H]
	\centering
	\includegraphics[width=0.8\columnwidth]{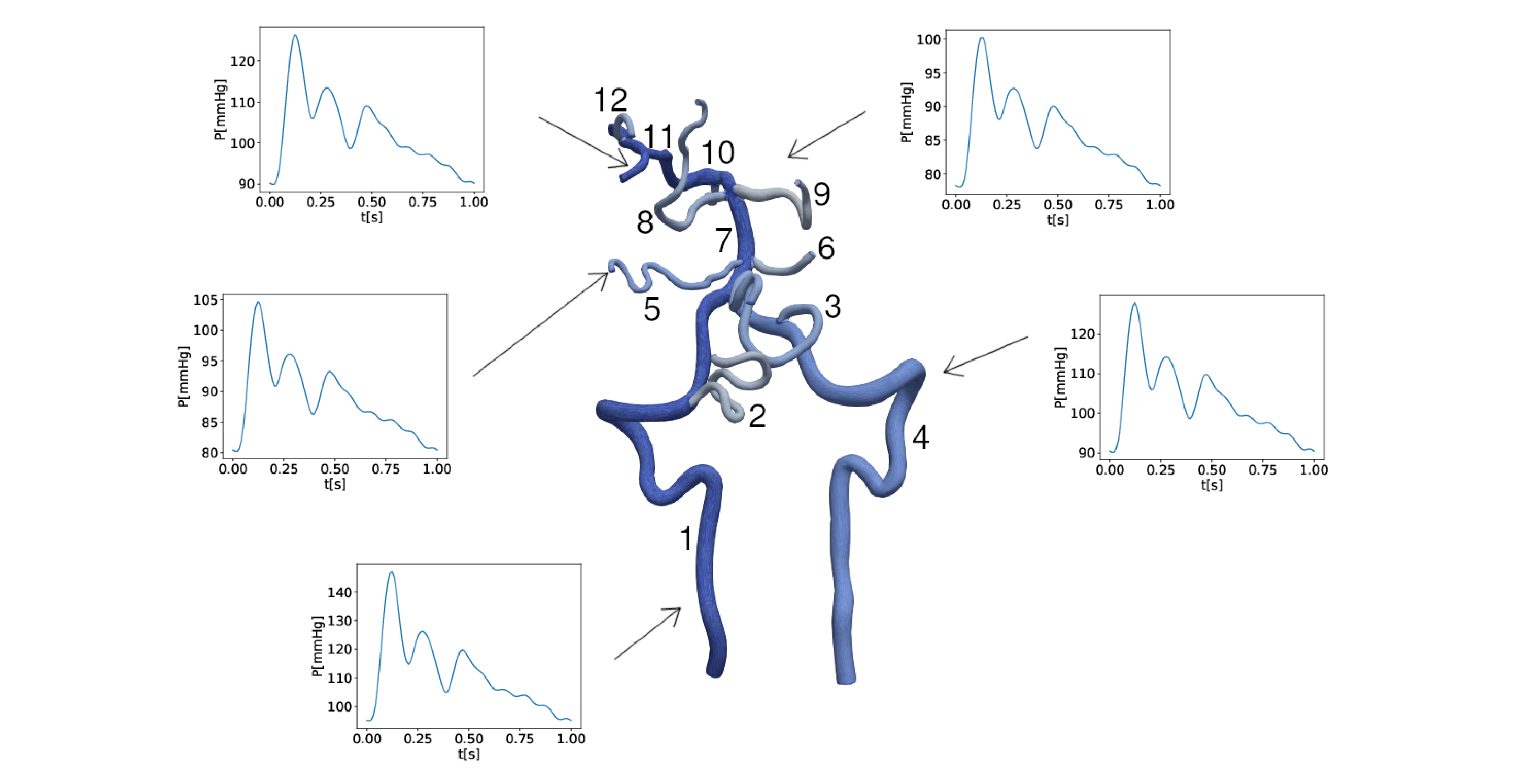}
	\caption{We present the jaxFlowSim results for different arteries of the Cerebral anatomy.}
	\label{fig:cerebral}
\end{figure}

The results of the abdominal aorta model are presented in Figure \ref{fig:abdominal} and the corresponding list of arteries is presented in Table \ref{tab:abdo}. The final example from the VMR database is a cerebral anatomy. The results of the proposed methodology are shown in Figure \ref{fig:cerebral}. The corresponding table containing the vessel indices and label is presented in Table \ref{tab:cere}. 
\begin{table}[h!]
	\begin{center}
		\begin{tabular}{|c|l|}
			\hline
			Legend & Vessel Name\\
			\hline
			$\mathbf{1}$& aorta \\ 
			$\mathbf{2}$& celiac trunk \\
			$\mathbf{3}$& celiac branch \\
			$\mathbf{4}$& superior mesentric artery \\
			$\mathbf{5}$& left renal artery \\
			$\mathbf{6}$& right renal artery \\
			$\mathbf{7}$& left common iliac artery \\
			$\mathbf{8}$& left internal iliac artery \\
			$\mathbf{9}$& right common iliac artery \\
			$\mathbf{10}$& right internal iliac artery \\
			\hline
		\end{tabular}
	\end{center}
	\caption{We list the 10 arteries of the abdominal model displayed in Figure \ref{fig:abdominal}.}
	\label{tab:abdo}
\end{table}

\begin{table}[h!]
	\begin{center}
	\begin{tabular}{|c|l|}
			\hline
			Legend & Vessel Name\\
			\hline
			$\mathbf{1}$& right vertebral artery \\ 
			$\mathbf{2}$& right posterior meningeal branch of vertebral artery \\
			$\mathbf{3}$& left posterior meningeal branch of vertebral artery\\
			$\mathbf{4}$& left vertebral artery \\ 
			$\mathbf{5}$& right anterior inferior cerebellar artery \\
			$\mathbf{6}$& left anterior inferior cerebellar artery \\
			$\mathbf{7}$& basilar artery \\
			$\mathbf{8}$& right superior cerebellar artery \\
			$\mathbf{9}$& left superior cerebellar artery \\
			$\mathbf{10}$& right posterior central artery \\
			$\mathbf{11}$& right posterior cerebellar artery \\
			$\mathbf{12}$& right posterior communicating artery \\
			\hline
		\end{tabular}
	\end{center}
	\caption{We list the indices and labels corresponding to the vessels displayed in Figure \ref{fig:cerebral}.}
	\label{tab:cere}
\end{table}

\begin{figure} [!ht]
	\centering
	\includegraphics[width=0.8\columnwidth]{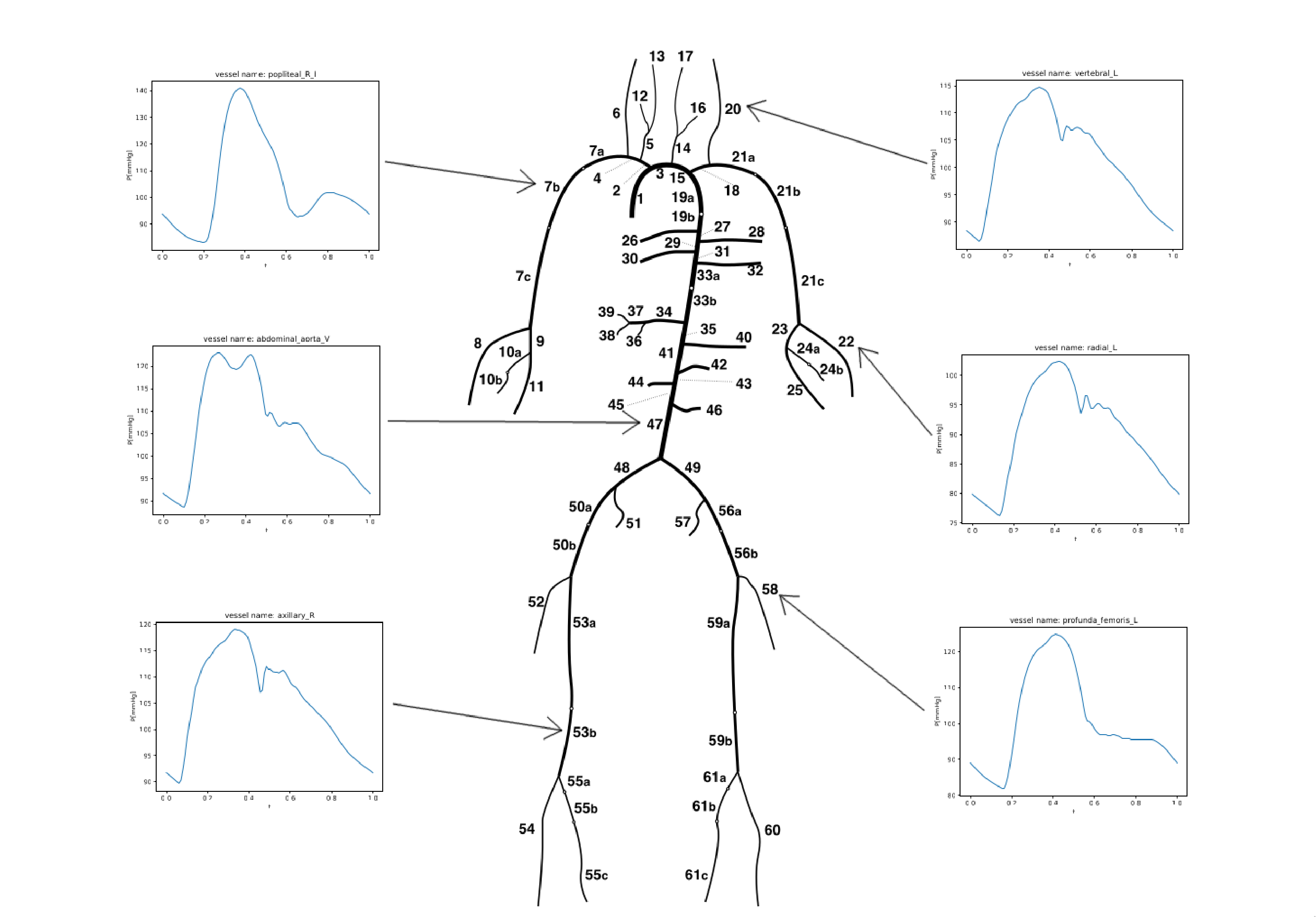}
	\caption{We present the jaxFlowSim results for different arteries of the ADAN56 anatomy. This sketch is adjusted from Boil et. el. \cite{boil2015}.}
	\label{fig:adan56}
\end{figure}

The Adan56 anatomy contains 56 vessels and is by far the largest vascular network considered in this manuscript. In Figure \ref{fig:adan56} the pressure waveforms for different vessels of the Adan56 geometry are presented. 
 We observe a stark difference between the pressure waveforms resulting when considering the anatomies taken from the VMR library and the ADAN56 network. This is the amount of variability between the form of different pulse waves corresponding to different arteries in the ADAN56 network as opposed to the VMR anatomies. This is probably a result of the way the Windkessel or RCR parameters were set to simulate the pulse-wave propagation problem in the different vascular networks. For ADAN56, the resistances and the compliance were carefully calibrated by Blanco et al. \cite{blanco2014anatomically,blanco2014blood} so that the 1D model provides physiologically realistic predictions. Because RCR parameters are not provided for all anatomies in the VMR library, they were computed using a common approximation as described in \ref{sec:ing}. The more physiologically accurate Navier-Stokes predictions for the meticulously assigned parameters in the ADAN56 model, as opposed to the approximated parameters assigned for the VMR anatomies, show the importance of calibrating the Windkessel parameters. This issue can also be made apparent by a closer examination of the different calibrating techniques considered for the RCR parameters presented in K{\"o}ppl et. al. \cite{köppl2023dimension}. The differentiable simulation framework proposed in this work is a step towards a more efficient solver calibration, as it allows advanced parameter inference techniques that consider gradient computations to be used.

\subsection{Scaling to Different Anatomies} \label{sec:sc}
\begin{figure} [!ht]
	\centering
	\includegraphics[width=0.5\columnwidth]{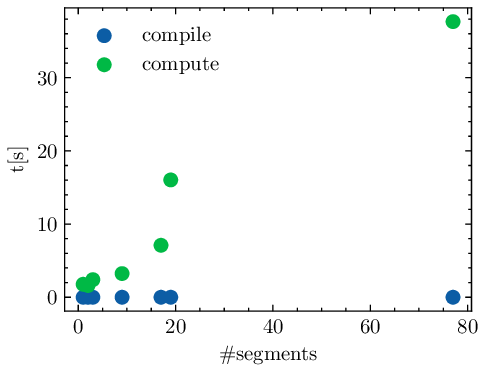}
	\caption{We present the compute and compile times of the jaxFlowSim solver for different number of vessels. We observe that the total compute times scales almost linearly with  the number of vessels.}
	\label{fig:timing}
\end{figure}

In Section \ref{sec:jax} of the Appendix, it is explained that the JAX interpreter performs an optimization of the code execution that results in a significant overhead in the overall wall-clock time. This begs the question of whether code optimization dominates overall time and how the optimization process scales for a growing number of segments. If the compilation time does not increase linearly or exponentially with the number of vessels, the compilation overhead is worth having for large anatomies. The compilation and computation time are presented against the number of segmented vessels in Figure \ref{fig:timing}. 
The compile time remains almost constant, around $10^{-4}$ seconds, even when the number of vessel segments increases approximately $80$ times. This results by avoiding built-in Python loops in favor of JAX loop constructions; thus the compile time is kept more or less static since these kinds of loop do not have to be unrolled during compilation; see Section \ref{sec:al} of the appendix for more details. Hence, for larger network simulations, there exists barely any additional computation involved. In general, the compile time is negligible even for simulations on small numbers of vessels. It can also be observed that the computation times obey an almost linear scaling as a function of the number of vessel segments and considering the simulation for the 19 vessel geometry as an outlier. The convergence of the solution in a simulation is heavily dependent on the specific geometry and conditions considered. Therefore, it is reasonable to assume that this kind of behavior might appear when considering scaling laws for simulations with different underlying conditions. It is important to mention that the comparison is performed for code execution on a CPU. The JAX language is device-agnostic, which means that \emph{the same exact code runs on a GPU or a TPU device}. The reason we make this comparison on CPUs is that the anatomies we consider are small, and the massive parallelization capabilities of GPUs cannot be properly leveraged.  However, for larger anatomies and scenarios that require running multiple simulations in parallel, such as sensitivity analysis, executing the solver on a GPU device would be advantageous.

\subsection{Parameter Inference} \label{sec:pi}
\begin{figure}[h!]
\centering
\includegraphics[width=0.4\columnwidth]{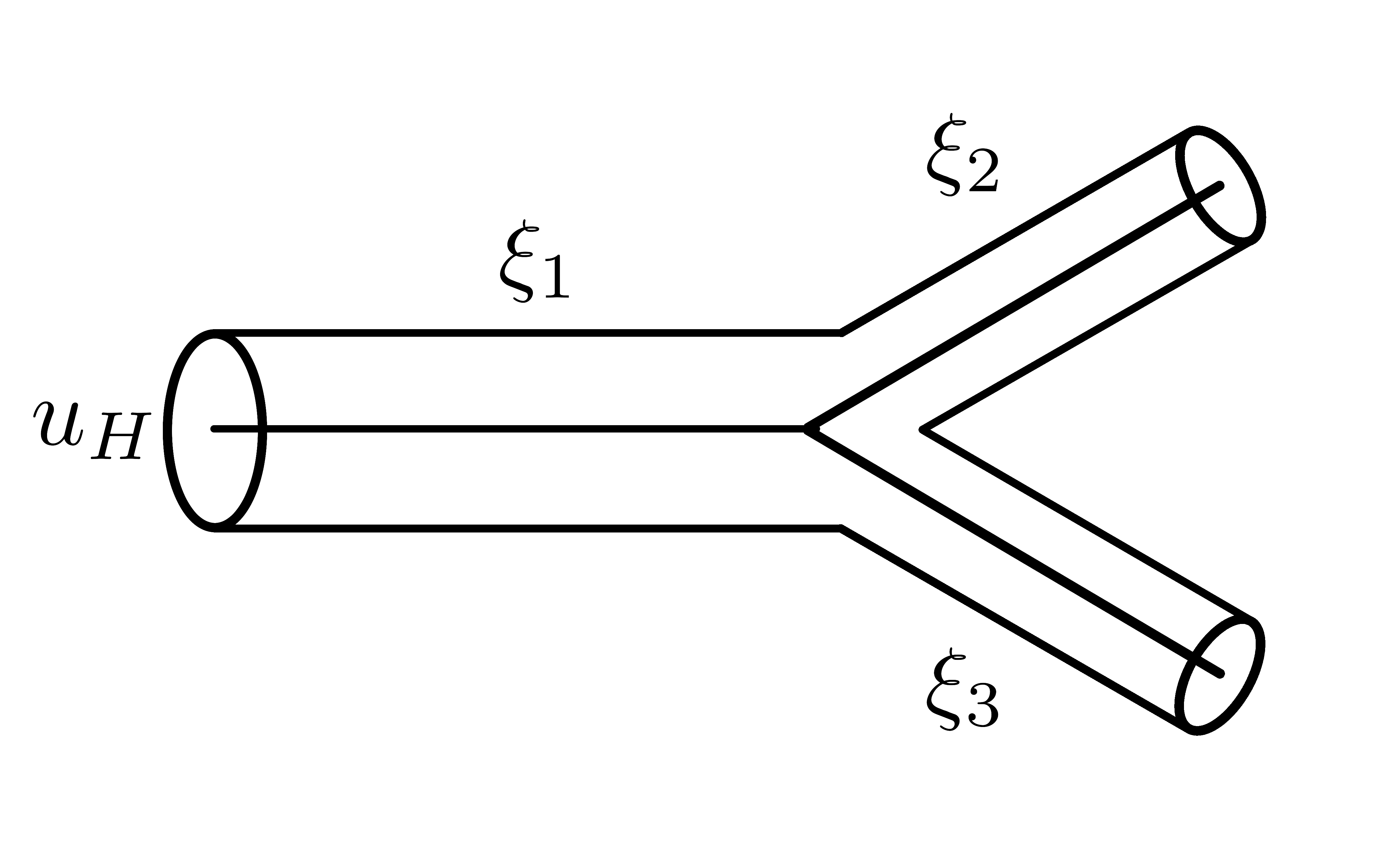}
\caption{A sketch of the geometry considered for the parameter inference.}
\label{fig:bifurcation}
\end{figure}

In this section, we employ the proposed differentiable solver to infer the parameters of a simple geometry containing one bifurcation, as shown in Figure \ref{fig:bifurcation}. The simulation requires a set of parameters and conditions to be decided, which means $\xi = \{\xi_1, \xi_2, \xi_3 \}$. We define $\xi_i = \{l_i, \beta_i, A_0^i, R_i, C_i\}$, where $l_i$ the length of the vessel, $A_0^i$ the equilibrium cross-sectional area, $R_i$ the vascular resistance, and $C_i$ the vascular compliance, if the vessel has an outlet at its right end. We consider an initial value for the velocity $u_0 = 0$ and a boundary value on the left $u(0, t) = u_H$. Calibrating the solver to specific conditions translates into identifying the parameters in the set $\xi_i$, such as $R_i, C_i$ that cannot be measured noninvasively from the data. We construct two parameter inference pipelines to test the capabilities of the solver, a deterministic and a probabilistic.

\subsubsection{Deterministic Inference}
We assume the relation
\begin{equation*}
    \mathbf{P}_\xi = \mathcal{H}(\xi),
\end{equation*}
where $\mathcal{H}$ is the PDE operator mapping the parameter set $\xi$ to the solution for the pressure $\mathbf{P}_\xi$.
We also introduce the pressure values produced by our solver, when given the parameter $\xi$, as 
\begin{equation*}
    \{ \mathbf{y}^{(n,m)}_\xi \}_{n,m=1}^{N,M}, N,M \in \mathbb{N}
\end{equation*} 
at the the $n$-th time-step $t_n$ and the $m$-th location $x_m$ equally distributed throughout the bifurcation system depicted in \ref{fig:bifurcation}. We note the obvious relation between the result of the PDE operator and our solvers results
\begin{equation*}
    \mathbf{y}^{(n,m)}_\xi \approx \mathbf{P}(t_n, x_m)
\end{equation*} 
where the approximate equality comes from numerical approximations and floating point arithmetic limitations. The objective is find $\bar{\xi}$, so that
\begin{equation*}
    \bar{\xi} = \underset{\xi \in \Xi}{\arg\min} L(\xi)
\end{equation*}
where 
\begin{equation*}
    L(\xi) = \frac{1}{M} \sum_{m=1}^{M} \frac{\sum_{n=1}^{N} \left( \mathbf{y}^{(i,j)}_\xi - \hat{\mathbf{y}}^{(i,j)} \right)^2}{\sum_{n=1}^{N} \left( \mathbf{y}^{(i,j)}_\xi \right)^2} \label{eq:loss}
\end{equation*}
and $\hat{\mathbf{y}}$ is the quantity of interest that corresponds to parameter $\hat{\xi}$. The loss is minimized by applying a gradient descent strategy which in simplified terms can be described as
\begin{equation*}
    \xi^{k+1} = \xi^{k} - \lambda F \left( \nabla_{\xi} L \left( \xi^k \right) , \nabla_{\xi} L \left( \xi^{k-1} \right) , ..., \nabla_{\xi} L \left( \xi^{0} \right) \right) , k \in \mathbb{N} \cup \{0\}, \xi^k = \tilde{\xi}
\end{equation*}
where $\lambda$ is a so-called learning rate, $F$ is a function compounding information from any previous optimization step, the gradients $\nabla_{\xi} L$ are computed by performing automatic differentiation through the solver, and $\tilde{\xi}$ is the initial value set for the optimization process. To increase the efficiency of the optimization process, we simulated for  time-steps despite our simulation not having converged yet. However, we could still observe rapid convergence of the optimization process with the expected results. The gradient optimization process was performed using the Adafactor optimizer \cite{shazeer2018adafactor}, as implemented through the Optax library \cite{deepmind2020deepmind}. We present here two results of the optimization process described previously. We first consider the case where the parameter set is only the terminal resistance of the first output vessel (denoted by $\xi_2$ in \ref{fig:bifurcation}) $\xi = \{R_{1,1}\}$. Then we will present a second case, where the parameter set is both the terminal and capillary resistances of both outlet vessels ($\xi_2$ and $\xi_3 $in \ref{fig:bifurcation}) $\xi = \{R_{1,1}, R_{2,1}$, $R_{1,2}, R_{2,2}\}$. 

\begin{figure}[h!]
\centering
\includegraphics[width=0.4\columnwidth]{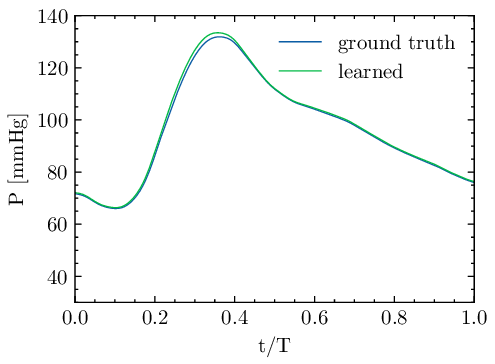}
\includegraphics[width=0.4\columnwidth]{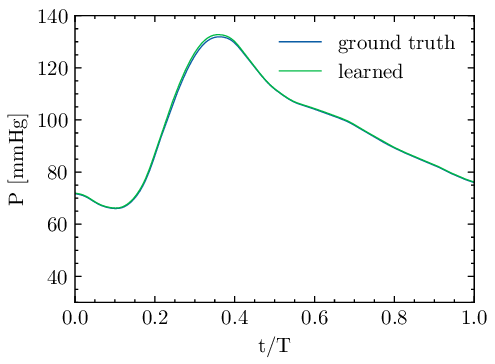}
\includegraphics[width=0.4\columnwidth]{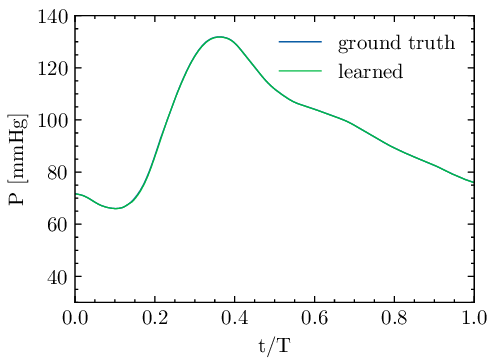}
\includegraphics[width=0.4\columnwidth]{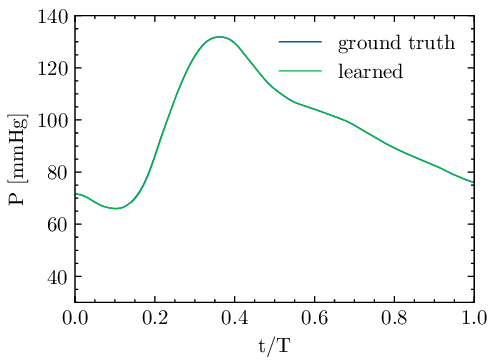}
\caption{Optimization of the $R_{1,1}$ parameter with $0, 100, 500, and 1000$ iterations from top left to bottom right. The Adafactor optimization already shows improvement after 100 iterations. Hence, the ground truth and the learned pressure wave overlap. }
\label{fig:optax_1}
\end{figure}

The pressure waves computed in the middle of the first outlet vessel during the optimization for the first case can be seen in Figure \ref{fig:optax_1}. We set $N=1000$, implying that all the time steps that we simulated were used in our loss function. The measurements were evaluated at $M=2$ locations at the end of each outlet vessel, respectively. The parameter set with which the ground truth was computed is $\hat{\xi} = \{6.8123\times 10^7\}$ and the optimization process was initialized with the parameter set $\xi^0=\{2 \times \log(1 + e^{s^0}) \times \hat{\xi} \}$, where the term $sp(s) :=\log(1 + e^s)$ is the softplus function of $s$ that was used in order to avoid negative, non-physical parameters. The scaling factor $s^0$ was drawn from a uniform distribution $U(0,2)$ resulting in $s^0=\{0.27158\}$ for the optimization process presented here. The optimization goal accordingly is to find $s$ s.t. $sp(s) = \{ 0.5 \}$. The Adafactor optimizer was run with a learning rate of $\lambda=10^{-1}$.
The corresponding losses and wall-clock times are recorded in \ref{tab:optax_1}.
\begin{table}[h!]
    \centering
    \begin{tabular}{cccc}
    \hline
         $\#$ Iterations & Loss $L$ & Wall-clock Time $[s]$ & scaling factor$s$\\
    \hline
        0 & -5.77005 & 0.0 & 0.83812 \\
    \hline
        100 & -6.66610 & 24.87188 & 0.68721\\
    \hline
        500 & -10.24914 & 56.47202  & 0.47333\\
    \hline
        1000 & -10.24913 & 96.01073  & 0.47333 \\
    \hline
    \end{tabular}
    \caption{For the optimization of the $R_{1,1}$ parameter, near identical waveforms can already be observed after 100 iterations and optimization steps after 500 iterations don't provide great improvement.}
    \label{tab:optax_1}
\end{table}
As can be seen from the plots, already after only 100 iterations the learned and the ground-truth pressure waves are barely distinguishable. Furthermore, when considering the values from the table, the loss quickly reaches a very low value and cannot be seen to improve after 500 iterations. After 500 iterations, we see that the value of the softplus factor $sp(s)$ is very close to the real value of $0.5$. We note that this result was achieved in less than one minute as can be seen in Table \ref{tab:optax_1}.

\begin{figure}[h!]
\centering
\includegraphics[width=0.4\columnwidth]{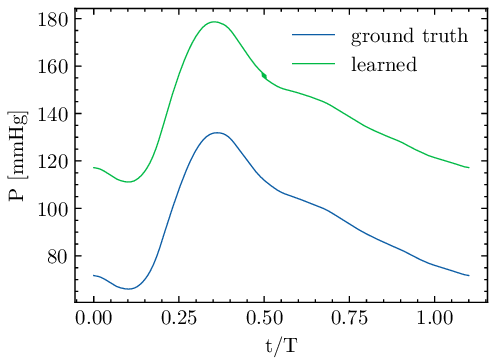}
\includegraphics[width=0.4\columnwidth]{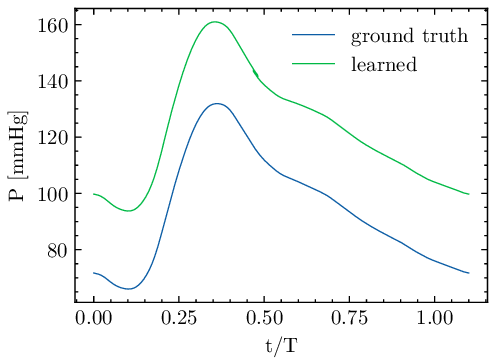}
\includegraphics[width=0.4\columnwidth]{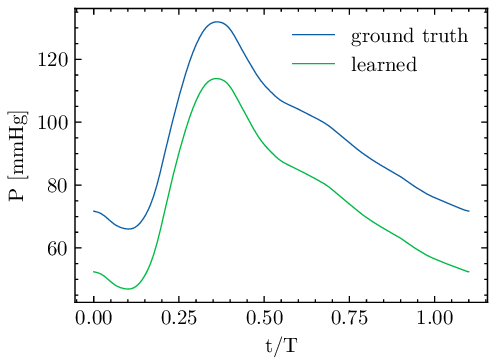}
\includegraphics[width=0.4\columnwidth]{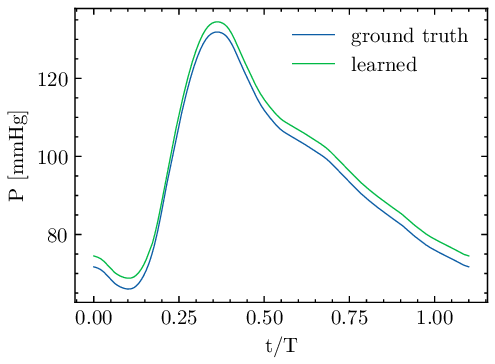}
\caption{Optimization of the $R_{1,1}, R_{2,1}, R_{1,2}, R_{2,2}$ parameter with $0, 100, 500, and 1000$ iterations from top left to bottom right. The Adafactor optimization reaches a high accuracy after 1000 iterations.}
\label{fig:optax_4}
\end{figure}

The pressure waves computed in the middle of the first outlet vessel during the optimization for the second case can be seen in Figure \ref{fig:optax_4}. We set $N=1$, implying that only the last time step that we simulated is used in our loss function. The measurements were evaluated at $M=2$ locations at the end of each outlet vessel, respectively. The parameter set with which the ground truth was calculated is $\hat{\xi} = \{6.8123\times 10^7, 3.1013\times 10^9, 6.8123\times 10^7\, 3.1013\times 10^9 \}$ and the optimization process was initialized with the parameter set $\xi^0=\{2 \times \log(1 + e^{s^0}) \times \hat{\xi}$, the scaling factor $s^0$ was drawn from a uniform distribution $U(0,1)$, resulting in $s^0=\{0.32333, 0.019362, 0.39539, 0.2471983 \}$ for the optimization process presented here. The optimization goal accordingly is to find $s$ s.t. $sp(s) = \{ 0.5, 0.5, 0.5, 0.5 \}$. The Adafactor optimizer was run with a learning rate of $\lambda=10^{-2}$.
The corresponding losses and wall-clock times are recorded in \ref{tab:optax_4}.
\begin{table}[h!]
    \centering
    \begin{tabular}{cccc}
    \hline
         $\#$ Iterations & Loss $L$ & Wall-clock Time $[s]$ & softplus $sp(s)$\\
    \hline
        0 & -4.62407 & 0.0 & $\{ 0.86782, 0.70287, 0.91026, 0.82436 \} $\\
    \hline
        100 &  -5.03054 & 25.40251 & $\{ 0.75823, 0.60453, 0.79681, 0.71454 \} $\\
    \hline
        500 & -12.09820 & 56.91697  & $\{ 0.51004, 0.36005, 0.50985, 0.43749 \} $\\
    \hline
        1000 & -13.40728 & 95.59291  & $\{ 0.50004, 0.52819, 0.49384, 0.50321 \} $ \\
    \hline
    \end{tabular}
    \caption{For the optimization of the $R_{1,1}, R_{2,1}, R_{1,2}, R_{2,2}$ parameter convergence is still occurring after 500 iterations.}
    \label{tab:optax_4}
\end{table}
As can be seen from the plots, the learned waveform first overshoots the groundtruth at 500 iterations but then converges close-by at 1000 iterations. Furthermore, when considering the values of the table, we see that after 1000 iterations, the value of the softplus factor $sp(s)$ is very close to the real value of $\{ 0.5, 0.5, 0.5, 0.5 \}$. We note that this result was achieved in less than two minutes as can be seen in \ref{tab:optax_4}.

\subsubsection{Probabilistic Inference} \label{par:stat_inf}

For the probabilistic inference case, a Hamiltonian Monte Carlo (HMC) approach is considered implemented through the NumPyro library \cite{phan2019composable}.  As in the deterministic inference section, we set the number of time steps that we simulate to $N=1000$. We infer the capillary and terminal resistances $\xi = \{R_{1,1}, R_{2,1}$, $R_{1,2}, R_{2,2}\}$ of both outlet vessels $\xi_2, \xi_3$ as shown in \ref{fig:bifurcation}. The pressure waves computed in the middle of the first outlet vessel $\xi_2$ during optimization can be seen in \ref{fig:optax_4}. We set $N=1000$, implying that only the last time step that we simulated is used in our loss function. Measurements were evaluated in $M=15$ locations evenly distributed throughout the 3 vessels $\xi_1, \xi_2, \xi_3$. The prior is set as $\xi^0=\{2 \times \log(1 + e^{s^0}) \times \hat{\xi}$ with the factor $\hat{\xi} = \{6.8123\times 10^7, 3.1013\times 10^9, 6.8123\times 10^7\, 3.1013\times 10^9 \}$. The prior of the scaling factor $s^0$ is distributed as $s^0 \sim \{N(-0.5,0.1), N(-0.5,0.1), N(-0.5,0.1), N(-0.5,0.1) \}$. The optimization goal accordingly is to learn the posterior of $s$ s.t. $sp(\mathbb{E}[s]) = \{ 0.5, 0.5, 0.5, 0.5 \}$.
\begin{figure}[h!]

\centering
\includegraphics[width=0.4\columnwidth]{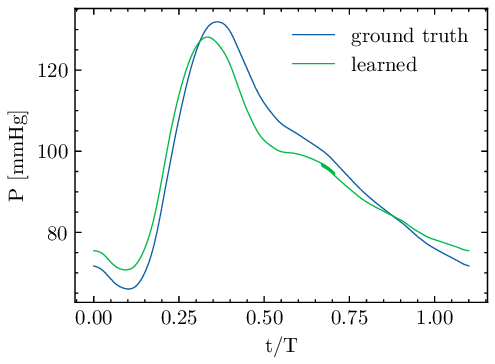}
\includegraphics[width=0.4\columnwidth]{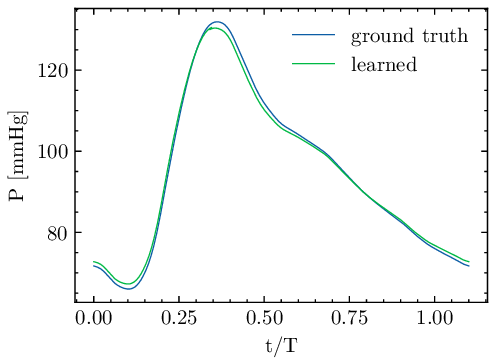}
\includegraphics[width=0.4\columnwidth]{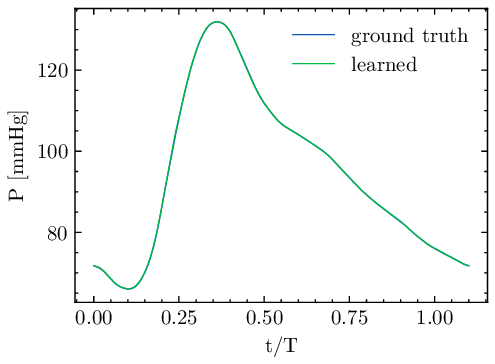}
\includegraphics[width=0.4\columnwidth]{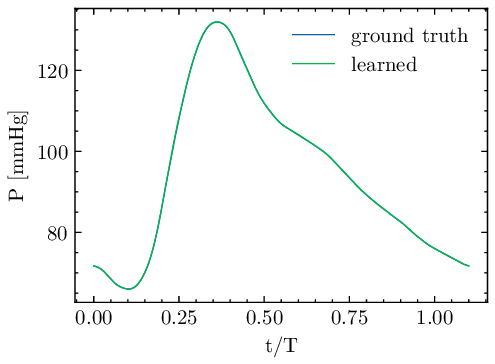}
\caption{Optimization of the $R_{1,1}, R_{2,1}, R_{1,2}, R_{2,2}$ parameter with $10$ warm-up and $10, 100, 500, 1000$ sampling steps from top left to bottom right. The Numpyro optimization reaches a high accuracy after 500 iterations.}
\label{fig:numpyro_4}
\end{figure}
The corresponding losses and wall-clock times are recorded in \ref{tab:numpyro_4}.
\begin{table}[h!]
    \centering
    \begin{tabular}{cccc}
    \hline
         $\#$ Sampling steps & Loss $L$ & Wall-clock Time $[s]$ & softplus $sp(\mathbb{E}[s])$\\
    \hline
        10 & -3.49109 & 54.06806 & $\{ 1.50797, 1.86507, 0.96325, 0.59960 \} $\\
    \hline
        100 &  -8.00814 & 77.98726 & $\{ 0.53993, 0.43278, 0.55722, 0.68136 \} $\\
    \hline
        500 & -9.69493 & 209.45971  & $\{ 0.50847, 0.46289, 0.52221, 0.54313 \} $\\
    \hline
        1000 & -10.51279 & 367.41249  & $\{ 0.51037, 0.48446, 0.51104, 0.51572 \} $ \\
    \hline
    \end{tabular}
    \caption{For the optimization of the $R_{1,1}, R_{2,1}, R_{1,2}, R_{2,2}$ the waveform is already close to identical after $500$ sampling steps but the loss function keeps decreasing slightly up until $1000$ sampling steps. }
    \label{tab:numpyro_4}
\end{table}
As can be seen in the plots, the learned waveform coincides with the ground truth after $500$ sampling steps. This is achieved in less than 4 minutes with a relatively small loss function. Furthermore, we observe that after $500$ sample steps the values of $sp(\mathbb{E}(s))$ are very close to the baseline of $\{ 0.5, 0.5, 0.5, 0.5 \}$, as shown in Table \ref{tab:numpyro_4}.

\section{Discussion} \label{sec:conc}
In this paper we have proposed a differentiable 1D-hemodynamics solver for both simulating pulse-wave propagation in complex anatomies and performing parameter inference. The underlying motivation of such simulations is the need for cardiovascular simulations in medicine, specifically in personalized medicine. The aim of a differentiable solver is to enable efficient and fast calibration to unknown patient flow conditions from data. The construction of the solver was achieved by first reducing the 3D-NSE to their 1D equivalent and defining the MUSCL numerical method for the system of equations. The differentiability property results from the implementation of the solver in the JAX language. We validated the proposed solver against the openBF solver and show the benefits in computational efficiency as well as the linear scaling of the wall-clock time with respect to the number of vessels. Moreover, we show that the time it takes for the JAX interpreter to optimize the code execution is near constant over different anatomies of different sizes. Furthermore, we showed that our solver can produce physiologically realistic results on models extracted from the Vascular Model Repository.  Moreover, we demonstrated the feasibility of performing both probabilistic and deterministic parameter inference on a geometry containing a single bifurcation. Although the main goal of making a differentiable 1D-hemodynamics solver was achieved, there are still improvements that need to be performed. First, the execution times on GPUs need to be improved, especially the treatment of junctions. Solving for junctions involves many small linear systems, a process that is not readily parallelizable in GPUs. Reformulating this part of the solver, perhaps in the form of one large sparse system, may lend itself better to GPU execution. Despite this being a seemingly inefficient way of solving the problem, considering how GPU computing works and combined with the optimizations from JAX, this might lead to better performance. It is also possible that the advantages of computing on a GPU could only be leveraged for very large systems or when running many simulations simultaneously. Furthermore, we provided a differentiable solver which enables more efficient parameter inference. However, to make this process useful for realistic applications, we need to perform an extensive test with larger vascular networks and a number of parameters to define the extent to which this process can be performed accurately. We consider the parameters of the Windkessel model to be a favorable target for such future work since they depend heavily on many physiological factors and there is no straightforward method of determining them for a given vascular network. The differentiable solver could be also used to for a solver-in-a-loop approach where the laws governing the behavior of the system can be substituted with parametric functions whose parameters are found using data. This is especially useful for laws that are based on empirical relations, which leads to there being no conclusive correct way to model something within the simulator. A prime example for this is the tube law, provided in Equation\ref{eq:p_tot}. Although there is a great deal of literature suggesting and comparing different methods of computing the tube law, there is no method that is considered to be the best or even fundamentally correct \cite{gomez2017analysis}. It would therefore be interesting to use data to infer this law and compare the results to the values obtained by using one of the given models from the literature.

\section*{Acknowledges}
The authors thank Siddhartha Mishra for his valuable input to the manuscript and the valuable discussions throughout the progress of this work. G.K. acknowledges support from Asuera Stiftung via the ETH Zurich Foundation. 

\bibliographystyle{unsrt}
\bibliography{references/references}

\appendix

\section{Implementation} \label{sec:impl}
Since differentiability was a priority in the implementation of this model, the JAX library is used. To maximize the full effectiveness of JAX, some special considerations need to be made. In this section, we provide an overview of the different functionalities of the JAX library and how they relate to the solver implementation.

\section{JAX} \label{sec:jax}
JAX used to stand for "Just After eXecution", which refers to the manner in which JAX used to optimize during compilation. JAX achieves optimization during compilation, among other things, through tracing. During tracing, the code is optimized with the help of a computational graph. Essentially, the code is parsed without executing it, in order to detect inefficiencies in data movement and computation. While JAX still uses tracing to speed up code execution, it is not done in a "Just After eXecution"-manner. Later on, JAX was thought of as a recursive acronym "JAX is Autograd and XLA" in the spirit of the well-known acronym GNU ("GNU is Not Unix"). Autograd is now a deprecated Python library that can automatically differentiate native Python and NumPy code. XLA stands for "Accelerated Linear Algebra", and it is an open-source compiler that optimizes models from commonly used frameworks such as PyTorch, Tensor Flow, and also JAX. XLA optimizes these models on different hardware platforms (CPUs, GPUs, TPUs, etc.). Often the optimizations done by XLA are directly integrated back into the projects from which the original models stem. XLA is also part of JAX, but while JAX used to be based on Autograd, this is not the case any more, and JAX achieves automatic differentiation through its own implementation thereof. Therefore JAX is not considered an acronym nowadays and JAX is just considered a library that implements the features of Autograd, in it's own way, while including XLA optimization. However, automatic differentiation and XLA optimization cannot be applied to any naive Python code; there are some caveats that must be considered when implementing numerical methods using JAX. We will be addressing these caveats in the next five subsections starting with the functional purity required when programming with JAX.

\subsection{Functional Purity}
All JAX code has to be written in a functionally pure manner, which means that the functions should not produce side effects. Moreover, all inputs to a function need to be passed explicitly through the function's arguments, while all outputs need to be returned explicitly through the return variables. Hence any kind of global variables should be avoided at all costs. The reason is that global variables can produce unwanted side effects. For example, if a global variable has a certain value at compile time and changes it's value during code execution, it will in the eyes of JAX, always keep the value it had assigned at compile time. The described behavior might lead to unexpected results, as a side effect of the compilation JAX performs. Furthermore, this behavior is not guaranteed. Therefore, writing a Jax implementation with fully reproducible behavior can only be achieved with functional purity.

\subsection{Vectorization} \label{sec:vect}
JAX's optimization is geared towards leveraging the capabilities of GPUs. This means that vectorization is key to making full use of JAX's optimization. The equations of the MUSCL scheme within a single vessel are very suitable for vectorization. This is the case because the numerical scheme describes an update formula for each point that can be executed in parallel. Vectorization can even be applied more widely using the techniques padding (Section \ref{sec:pdd}) and masking (Section \ref{sec:msk}) described next.

\subsection{Padding} \label{sec:pdd}
The MUSCL solver presented in Section \ref{sec:muscl} is defined for a single vessel. However, all of the MUSCL solvers applied to the individual vessels could be performed simultaneously and even in a vectorized manner. To do so, the values to be computed by the MUSCL solver of each vessel are all written into one single array, and one MUSCL solve is performed on the entire array. To avoid errors at the ends of a vessel, where the values stored from one vessel would influence the computed values for another vessel, one can apply padding. This translates to adding zero entries between arrays to avoid two neighboring vessels from influencing one another. The errors that occur from collecting all data in one array then only affect the padded entries which are ignored when reading out the final results. After each MUSCL solve, when boundary conditions are computed, the padded entries are set equal to the values of the edges of the adjacent vessels. The padding approach not only allows for efficient application of the MUSCL solver to all vessels at once but also allows the use of an inhomogeneous mesh on the vessels of the simulation. That means that every vessel can be discretized according to its length, which is very useful since vessels sizes can differ. The reason why padding allows for inhomogeneous meshes is described in the next section.

\subsection{Static Array Sizes and Avoiding Loops} \label{sec:al}
The sizes of arrays handled in JAX codes need to be known at the time of tracing because the compiler requires all array sizes at compile time to optimize the code execution. Furthermore, it is recommended to avoid using loops because they need to be unrolled completely during JAX's compilation. One can consider JAX built-in loop constructions to avoid such issues. However they have restrictions and one needs to satisfy a very strict signature when using them. Specifically, the arrays handled in each iteration of the loop need to be of the same fixed size. So, if padding was not performed to make the sizes equal, the MUSCL solver would have to be applied to each vessel separately by looping over all vessels. This naive approach would result in long compilation times for cardiovascular networks consisting of many vessels due to unrolling. Another option is to employ a native construction, the so-called JAX \emph{fori\_loop}, which would avoid the large overhead at the compile time. However, the restriction of requiring the same-sized arrays in every iteration would make the simulation either inherently inefficient or inherently inaccurate since very small vessels would have to be simulated with fine meshes or vice versa large vessels with coarse meshes. Therefore, padding not only allows for increased vectorization but also provides the ability to use an inhomogeneous mesh while keeping the compilation time relatively short at the same time. In some cases loops cannot be avoided by padding alone, and another technique called masking has to be applied, which we will describe in the next section.

\subsection{Masking} \label{sec:msk}
At certain points within the MUSCL solver, ghost cells need to be inserted at the edges of the vessels.
In order to avoid loops, one could directly insert these values using a technique called masking. To do this, one needs to store five arrays. First, the original array  that the ghost cells are inserted, padded with a zero on either end. Second, store twice the original array shifted forwards and backwards padded with two zeros at the beginning or the end respectively; and  store twice a boolean array containing the locations of either the ghost cells at the inlet or the outlet. The padded original array, the forward shifted array, and the boolean array with the locations of the ghost cells at the outlets can be combined with the JAX equivalent of the NumPy where-function to insert the outlet ghost cells. The equivalent process is applied then to get the original array.

\section{Code Structure} \label{sec:cs}
The solver structure is denoted in pseudo-code in Listing \ref{lst:pc}. The simulation routine contains the runSimulation function that receives arguments from the config\_filename and $J$, the maximum number of time steps before terminating a simulation if it has not converged.  
The simulation is initialized through a function that reads and sets the initial conditions described in Section \ref{ssec:initial_conditions} of \emph{<path/to/network\_configuration>.yml}. 
Specifically we set the initial values ($u(z;0)$, $A(z;0)$, $Q(z;0)$, for each vessel; the left boundary condition ($P(0;t)$ or $Q(0;t)$, at all inlets; and the right boundary condition parameters ($R_t$, $R_1$, $C$, $R_2$) for the outlet vessels; the mesh size ($\Delta x$) and stiffness parameters ($\beta$, $P_{ext}$ ) for each vessel; and global parameters ($\rho$, $T$, $C_{CFL}$, $N$, where $N$ stands for the number of vessels in the network). Then we generate the matrix $J \times N$ $P_t$ used to store the pressure values in the center of each vessel for one cardiac cycle. Thereafter a while-loop is initiated where an entire cardiac cycle of length $T$ is computed at each iteration. In this while-loop we first initialize the time $t$, the step count for a cardiac cycle, and store a copy of the current pressure values to check for convergence.

Each cardiac cycle is itself described by a while-loop that computes the next time-step using the $C_{CFL}$ parameter, then updates the boundary value, and executes the MUSCL solver for all vessels advancing $\mathbf{U}_i^{n}$ to $\mathbf{U}_i^{n+1}$ $\forall i \in \{1,...,M\}$. Then the process saves the pressure at the center of each vessel for each individual time step, and finally advances the time step and increases the counter. The loop ends if the maximum allowed step $J$ is reached. After each cardiac cycle computation, the pressure differences in each vessel are compared to the previous values, and if they are smaller than $0.1$ mmHg, the simulations are considered to be converged. This is in accordance with the comment made in Section \ref{ssec:initial_conditions} that steady state needs to be reached to mitigate the effects of imprecise setting of the initial conditions.

\begin{lstlisting}[language=Python, caption=The code structure of an entire simulation is given here in pseudocode. Each line is detailed throughout this section., label=lst:pc, escapechar=|]
def runSimulation(config_filename, J)
	config = loadConfig(config_filename) |\label{ln:init_start}|
	simulation_data = buildArterialNetwork(config) |\label{ln:init_end}|

	P_t = [0] |\label{ln:pt}|

	converged = False |\label{ln:whout1}|
	while not converged: |\label{ln:whout2}|
		t = 0 |\label{ln:t0}|
		i = 0 |\label{ln:i0}|
		P_t_temp = P_t |\label{ln:cp}|
		while t < T:
			dt = computeDt(simulation_data) |\label{ln:cfl}|
			simulation_data = setBoundaryValues(simulation_data, dt) |\label{ln:bv}|
			simulation_data = muscl(simulation_data, dt) |\label{ln:muscl}|
			P_t[i,:] = savePressure(simulation_data) |\label{ln:svp}|
			t = t + dt |\label{ln:updt}|
		i = i + 1 |\label{ln:updi}|
		if i >= J
			break
		converged = checkConv(P_t, P_t_temp) |\label{ln:conv}|
						\end{lstlisting}

Note that the computation of the boundary conditions requires a for loop.  This for loop is constructed using the JAX \emph{fori\_loop} as discussed in Section \ref{sec:al}. The same goes for the two while-loops and their JAX equivalent \emph{while\_loop}.  The need for loops has been completely avoided when executing the MUSCL solver on all vessels. This was done by applying padding and masking, allowing for better code optimization and leveraging GPU resources more effectively. Finally, what is denoted in the Listing Line \ref{lst:pc} by \emph{simulation\_data} consists of multiple 2D arrays that contain global constants, vessel constants, and simulation quantities. While this format is less human-readable, than for instance having a class that describes a vessel, it allows for more efficient data movement and to easily adhere to JAX's code optimization requirements. 

\end{document}